\documentclass{ws-ijmpd}
\usepackage[super,compress]{cite}

\usepackage{amsmath}
\usepackage{amsfonts}
\usepackage{mathrsfs}

\begin{document}

\markboth{Martin Bucher}
{Physics of CMB (2015-2025)}

%
\catchline{}{}{}{}{}
%

\title{UPDATE ON THE PHYSICS OF THE \\COSMIC MICROWAVE BACKGROUND (2015-2025)}

\author{Martin Bucher\footnote{bucher@apc.univ-paris7.fr}}

\address{Laboratoire Astroparticules et Cosmologie (APC), Universit\'e Paris Cité/CNRS\\ 10 rue Alice Domon, 75013 Paris, France}

\maketitle

\begin{history}
\received{Day Month Year}
\revised{Day Month Year}
\end{history}

\begin{abstract}
The chapter titled ``Physics of the Cosmic Microwave Background Anisotropy'' appeared exactly one decade
ago in the compilation \textit{One Hundred Years of General Relativity: From Genesis and Empirical Foundations to Gravitational Waves, Cosmology and Quantum Gravity}.
This earlier contribution summarized the state of the field then as well as the underlying cosmology and gravitational physics. This contribution reports on what new
has occurred in the field during the intervening time 2015-2025. We also review future prospects.
\end{abstract}


\ccode{PACS numbers:}


\section{Introduction: The CMB in a Nutshell}

Observations of the 
Cosmic Microwave Background (CMB) anisotropies in temperature and polarization provide us with a glimpse of the state of the Universe 
at approximately
$1/1100$ its present size at a time only some 380,000 years after
the putative Big Bang. The CMB offers a two-dimensional view in projection onto the 
celestial sphere, unlike three-dimensional
galaxy and other surveys, which owing to the additional information in the radial direction 
in principle would be much more information rich if
their interpretation were not subject to nonlinearities and \textit{ad hoc} 
modeling of astrophysical processes, which at least with the present state of the art 
cannot be modelled from first principles.

\begin{figure}
\begin{center}
\includegraphics[width=0.49\textwidth]%
{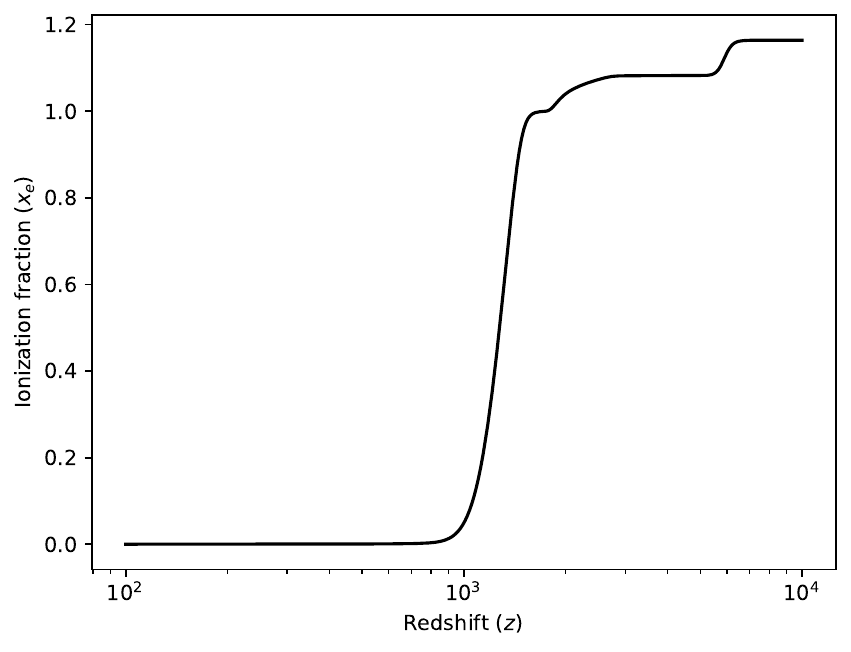}
\includegraphics[width=0.49\textwidth]%
{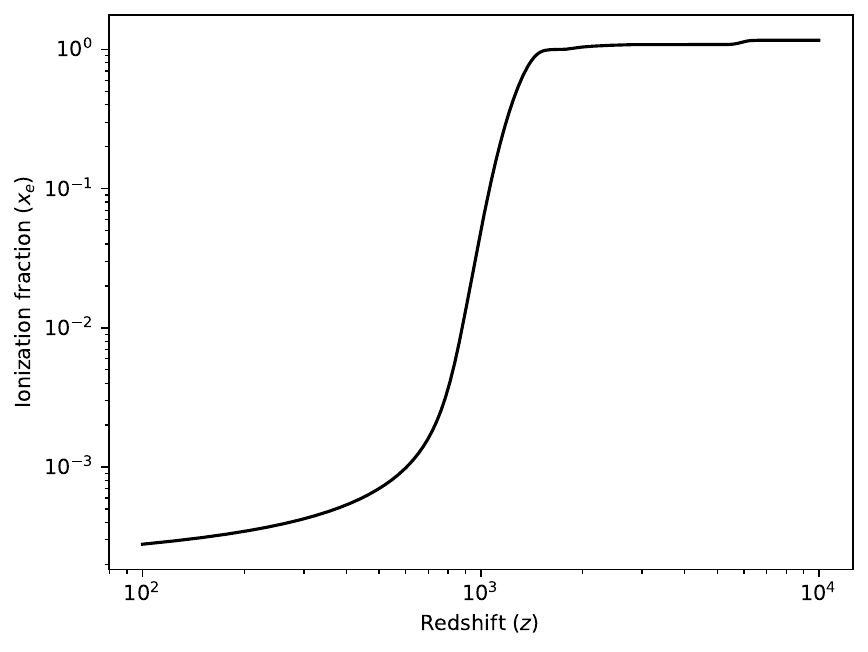}\\
\hbox to 0.175\textwidth{}(a) \hbox to 0.43\textwidth{}(b)\\
\includegraphics[width=0.49\textwidth]%
{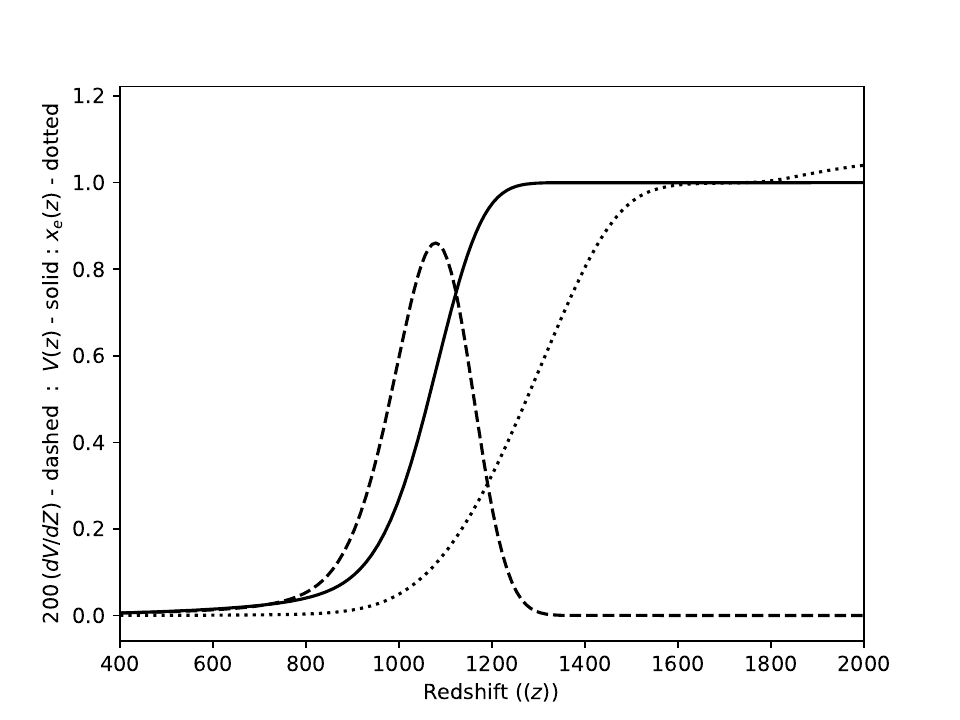}
\includegraphics[width=0.49\textwidth]%
{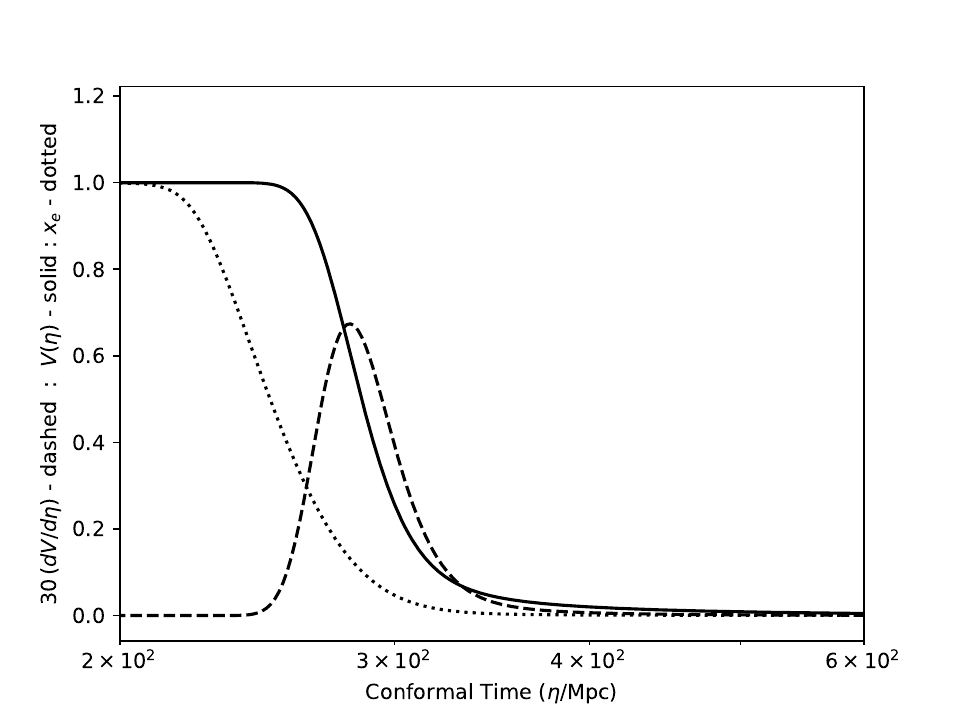}
\hbox to 0.175\textwidth{}(c) \hbox to 0.43\textwidth{}(d)
\end{center}
\caption{%
\textbf{Cosmological Recombination.}
As the universe expands and cools down, 
the `baryonic matter' (which in cosmological jargon also includes 
electrons in addition to ions and neutral atoms) transforms
from being a completely ionized plasma to being in an almost completely
neutral state. Panels (a) (linear vertical scale) and (b) (logarithmic vertical scale)
show how the electron ionization fraction $x_e$ evolves as a function
of redshift. Panels (c) and (d) show the optical depth and visibility
until today (with late-time reionization from stars and quasars turned off) 
as functions of redshift and conformal time, respectively.
Here the concordance six-parameter model is assumed. 
[Credit: Reprinted with permission from Ref.~\citen{martinBook}. Courtesy of Cambridge University Press.]}
\label{Fig:RecombFigure}
\end{figure}


The physics of how the CMB was imprinted is relatively simple and clean. It 
is believed to be well understood, at least if we exclude 
nonminimal scenarios introducing new physics. 
As the universe cooled down, the universe transformed from being
a fully ionized plasma to a gas of neutral atoms (mostly atomic
hydrogen but also some ${}^4\!H\!e$). During this early plasma phase, photons
scattering occurred frequently, so that photon directions are rapidly randomized.
The photons, ions, and electrons are tightly coupled, behaving
much like a fluid. When the contribution of the ions to the inertia
compared to the photon energy contribution is negligible, the speed of sound
is only slightly below $c/\sqrt{3},$ the sound speed
in an ultrarelativistic gas. As the universe expands and cools
down and as the universe starts to become matter dominated, the 
sound speed decreases resulting from the increasing 
inertia from the protons and helium ions. As the universe
expands and cools further, the formation of neutral hydrogen
and helium is favored, and the universe gradually becomes
neutral. Whereas the cross section for Thomson scattering by 
free electrons is large, the cross section for
the scattering of CMB photons by neutral atoms or molecules
is minuscule. This is because the CMB photon energy lies
well below the threshold for the excitation of an atom
from its lowest energy state. Thus 
the universe becomes transparent, causing
the photons to travel to us today with a very small probability
of being rescattered.\footnote{This is an oversimplification
because a small fraction of the CMB photons are rescattered due
to `reionization' caused by UV photons from the first 
generation of stars and quasars. This is a modest correction
discussed further below.} Some details of these processes are indicated in 
Figs.~\ref{Fig:RecombFigure}(a) and \ref{Fig:RecombFigure}(b), where the ionization fraction is plotted 
as a function of redshift. Figure~\ref{Fig:RecombFigure}(c) 
also plots the photon survival probability $V(z)=\exp (-\tau (z))$ 
and the associated visibility density $v(z)=(d/dz)V(z)$
where $\tau (z)$ is optical depth from redshift $z$ to us today (i.e., $z=0$), 
and Fig.~\ref{Fig:RecombFigure}(d)
shows the same as a function of conformal time
(normalized so that $a_0=1$ today).
The conformal 
time today in the standard cosmology is $\eta _0=14.4\,\textrm{Gpc},$
or $46.8\,\textrm{Gyr}$ when converted into a time.

We presently have precise maps of the CMB temperature
anisotropies close to the cosmic variance limit for those scales
dominated by the primordial signal. Observations from space (from the 
COBE, WMAP, and Planck space missions) provide the best data on large
angular scales. 
Ground-based observations
have better angular resolution because of larger telescope aperture sizes
and thus provide better data
on smaller angular scales. 
By combining the two, we may obtain 
the power spectrum close to the cosmic variance limit from $\ell =2$ to
$\ell \approx 4\times 10^3,$ beyond which other effects such as gravitational lensing
and point sources start to dominate.

Observations of the CMB polarization in both the $E$ and $B$ modes 
provide corroboration for the cosmological models deduced from 
CMB temperature data alone, possibly also combined with other 
non-CMB data. For pure scalar perturbations (the only ones detected to date), 
at linear order only $E$ mode polarization is predicted. The 
$E$ mode polarization anisotropies 
predicted
using the cosmological parameters deduced from the
temperature anisotropies are broadly consistent with the observed $E$ mode polarization
anisotropies (see Fig.~\ref{planckTEandEE}). 
$E$ mode polarization anisotropies do allow certain degeneracies to be 
broken, in particular the degeneracy between the recombination optical depth $\tau $
and the primordial amplitude of the scalar cosmological perturbations $A_S.$
The $B$ mode polarization anisotropy of the CMB
has been detected on small angular scales, and this signal is consistent
with gravitational lensing, which is a secondary CMB anisotropy and occurs
at quadratic order in the scalar perturbations. Of more fundamental interest
is the search for primordial $B$ modes, which arise from pure tensor modes, as discussed
further below. At present, we have only upper bounds on a possible nonzero
tensor mode amplitude. 

The book chapter ``Physics of the Cosmic Microwave Background Anisotropy'' 
\cite{bucherBookChapter} (finalized in October 2014 and 
also published in Int. J. Mod. Phys. D \cite{bucherBookChapter-IJMPversion}) 
emphasized the theoretical aspects of the subject. On the theoretical
front, there has not been so much progress. The progress that has taken
place has been more in the area of CMB observation, as emphasized
in this contribution. We also offer where needed intuitive explanations to the 
relevant theory, covered in more detail in the 2015\cite{bucherBookChapter-IJMPversion}
and 2017\cite{bucherBookChapter}
versions of the earlier book chapter.
A more detailed account may be found in my book.\cite{martinBook}  
(See also the Staggs et al.~review.\cite{staggs})

\section{A Pre-2015 History of CMB Observations}


The CMB was first discovered by Penzias and Wilson at Bell Labs in 1964, although 
it could potentially have been discovered much earlier using molecular thermometers
had the observations by McKellar and others been fully interpreted.
The observations by Penzias and Wilson received a lot of attention and provide
strong evidence for the hot Big Bang expanding universe model, although the 
defenders of steady state cosmological models did not immediately throw in the towel and
continued attempting to devise models that would explain a nearly isotropic perfect
blackbody background within the framework of a steady state universe mdoel.
In any case, the interest in establishing better limits on deviations
from isotropy in the temperature of this background and on possible
deviations from a perfect blackbody energy spectrum was immediately
apparent, and from 1964 through the early 1990s there was a long series
of experiments along these lines pushing the upper bounds lower and lower
(as well as a few false detections). But because the necessary technology was 
not then available, a first detection had to await the COBE DMR detection.
The pre-COBE story of CMB observations is treated in detail in 
Refs.~\citen{partridgeBook} and \citen{findingBigBang}. 

An anisotropy in the CMB temperature was first discovered in 1992 by the COBE 
(Cosmic Background Explorer) DMR (Differential
Microwave Radiometers) team led by George Smoot, for which he was awarded
the 2006 Physics Nobel Prize together with John Mather.\cite{cobeDMR}
John Mather led the COBE FIRAS 
(Far Infrared Absolute Spectrophotometer)
team, which measured the CMB energy spectrum 
with unprecedented precision, providing additional confirmation of the 
hot Big Bang expanding universe model and making it virtually impossible
to devise plausible, noncontrived steady state universe models.\cite{cobeFIRAS}

The COBE DMR experiment surveyed the entire sky in the three frequency bands 
31, 53, and 90 GHz (where the CMB anisotropy is expected to dominate) 
using pairs of corrugated microwave feedhorns pointing in directions 
on the sky separated by $60^\circ $ (at $\pm 30^\circ $ with respect to
the spin axis of the satellite). The amplifier input was rapidly switched
between the two horns and only the difference was recorded. This is because
the amplifier zero-point and gain are unstable on account of low-frequency
excess noise (also known as $1/f$ or flicker noise), so absolute measurements
cannot be taken at the necessary precision. Only differences 
between measurements closely separated in time provide meaningful information. 
The satellite rotated at 0.8 rpm about its spin axis, 
always directed away from the earth. Moreover every approximately $104$ minutes 
the satellite orbited the earth and the plane of the orbit
was situated at approximately a right angles to the Sun.
The temperature differences recorded allowed a full-sky map 
of the sky temperature to be reconstructed (with only one undetermined
mode, namely the zero point of the map). 

\begin{figure}
\begin{center}
\includegraphics[width=0.7\textwidth]{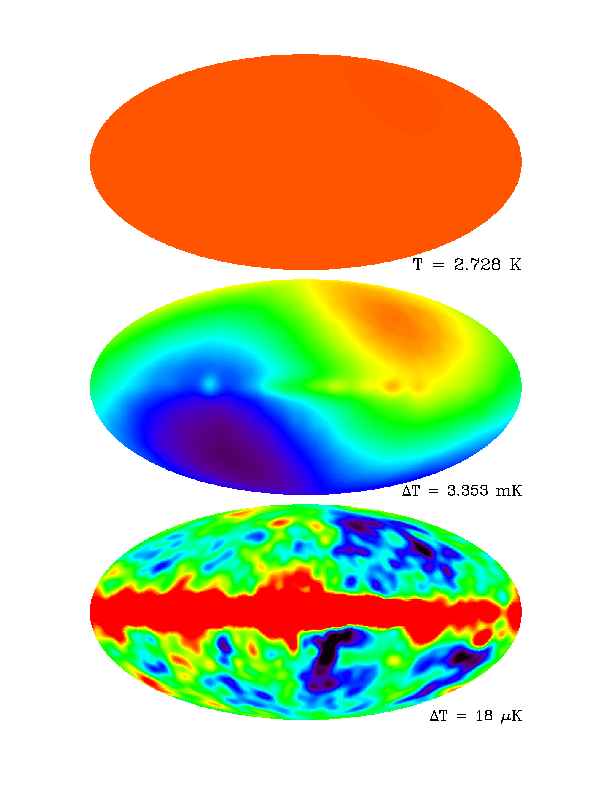}
\end{center}
\caption{%
\textbf{The Microwave Sky as Seen by the COBE DMR Instrument.}
The three panels show the microwave sky in Galactic
coordinates using a Mollweide projection. The Galactic
center is in the middle and the Galactic plane maps
onto the equator.
The top panel shows the microwave sky observed by the COBE
DMR instrument using a colorscale that includes $T=0,$
so that the map appears completely isotropic.
In the middle panel the monopole component (or map average) has
been subtracted, so that except for some blemishes near
the Galactic plane, the eye sees a pure dipole 
pattern, which can be attributed 
to our proper motion with respect to the cosmic rest
frame, although a subdominant intrinsic dipole 
contribution cannot be excluded. In the bottom panel both the 
monopole and dipole contributions have been subtracted,
so that one sees the primordial signal at high Galactic latitudes
and the Galactic emission around the Galactic equator.  
[Credit: NASA COBE Science Team. \copyright USG.]
}
\label{fig:COBEone}
\end{figure}

\begin{figure}
\begin{center}
\includegraphics[width=0.7\textwidth]{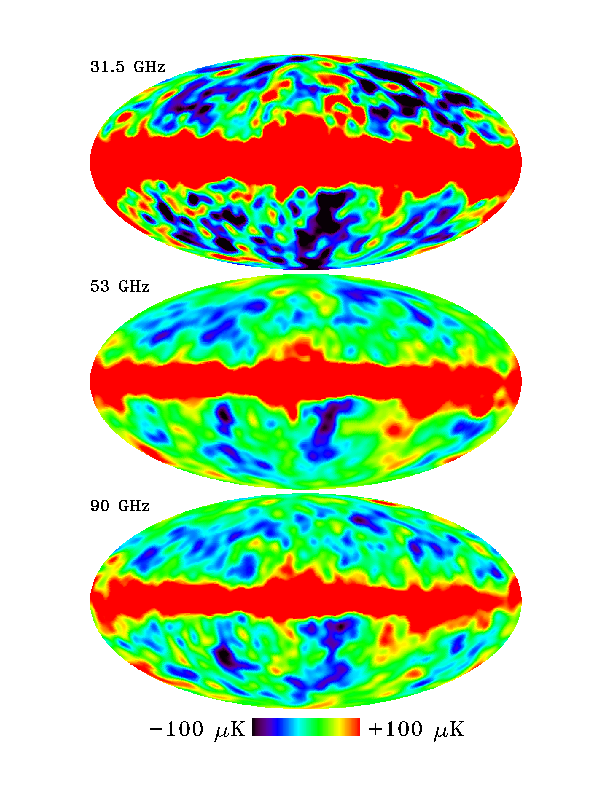}
\end{center}
\caption{%
\textbf{COBE DMR 31, 53 and 90 GHz Temperature Maps.}
[Credit: NASA COBE Science Team. \copyright USG.]
}
\label{fig:COBEtwo}
\end{figure}

Figure \ref{fig:COBEone} shows the microwave sky in Galactic coordinates as seen by 
the DMR instrument. Panel (a) 
uses a  colorscale that includes 0\,K and to the bare eye
no CMB anisotropy is visible.
In panel (b) the map average, or monopole harmonic,  
has been subtracted away, and except 
for a few blemishes around the Galactic plane, one sees what resembles 
a pure dipole anisotropy pattern, which can
be attributed to our motion relative to the CMB rest frame
due to the Doppler effect.
Finally
in panel (c) both CMB monopole and dipole have been
subtracted away, and except for a band around the Galactic plane (or equator), the spotted
pattern appears roughly isotropic. 

Figure \ref{fig:COBEtwo} shows  
the three COBE DMR frequency maps 
at 31, 53, and 90 GHz with the monopole and dipole subtracted away.
The band of excess emission about the Galactic plane is widest at the lowest frequency
and narrowest in the 90 GHz channel, but outside this band the emission appears
more or less independent of Galactic latitude. Perhaps most importantly, 
the patterns look more or less the same in the different frequency bands,
even though the 31 GHz band is substantially more contaminated than
the 90 GHz band by the emission from the Galaxy, which for the most part
arises from Galactic synchrotron emission, although there is also
a contribution from Galactic free-free (or Bremsstrahlung) emission. 

For comparing to theory, it is convenient to expand the 
sky temperature map into spherical harmonic components 
(exploiting the presumed statistical isotropy of the CMB), 
so that
\begin{equation}
\delta T(\hat {\boldsymbol{\Omega }})=
\sum _{\ell =0}^\infty 
\sum _{m=-\ell }^{+\ell }
a_{\ell m}^T~Y_{\ell m}^{\phantom{T}}(\hat {\boldsymbol{\Omega }}). 
\end{equation}
Here $\hat {\boldsymbol{\Omega }}=( \theta , \phi )$ indicates a direction on the celestial sphere.
We may define the `observed' angular power spectrum 
\begin{equation}
C_\ell ^{TT\, (\textrm{obs})}=
\frac{1}{(2\ell +1)}
\sum _{m=-\ell }^{+\ell }
a_{\ell m}^T~
a_{\ell m}^{T\,*},
\end{equation}
and the `theoretical' model power spectrum, defined as 
the expectation value 
\begin{equation}
C_\ell ^{TT\, (\textrm{th})}=
\left\langle 
a_{\ell m}^T~
a_{\ell m}^{T\,*}
\right\rangle   
\end{equation}
taken over a somewhat fictitious ensemble of an infinite number of independent realizations of our universe. 

Of course, science fiction aside, one can observe only one universe---namely our own. 
The experiment cannot be repeated, and this limitation 
is known as `cosmic variance.' If we know the true model (and  
beyond statistical isotropy also assume Gaussianity), each estimator 
\smash{$C_\ell ^{TT\, (\textrm{obs})}$} is an unbiased minimum variance estimator of 
the `true,' theoretical power spectrum 
\smash{$C_\ell ^{TT\, (\textrm{th})}$}, 
and \smash{$C_\ell ^{TT\, (\textrm{obs})}/C_\ell ^{TT\, (\textrm{th})}$} obeys a 
reduced $\chi ^2$-distribution 
with $(2\ell +1)$ degrees of freedom. The reduced $\chi ^2$-distribution (obtained by
dividing by its expectation value) has a variance of $2/(2\ell +1).$
 
If we coarse-grain the observed power spectrum at a resolution having a fractional bandwidth
of $\Delta \ell /\ell ,$ we are able to reconstruct the true model power spectrum with 
a fractional accuracy of 
\begin{equation}
\frac{\Delta C_\ell}{C_\ell}\approx 
\frac{1}{\ell }
\left(
\frac{\Delta \ell }{\ell }
\right) ^{1/2}
\label{CVeqnFirst}
\end{equation}
as can be derived using basic properties of the $\chi ^2$ distribution. 
A reasonable degree of smoothing might be to set $\Delta \ell $ to
be comparable to the scale of the acoustic peaks, or a more aggressive
smoothing of $\Delta \ell \sim \ell $ could be contemplated.\footnote{Actual
analysis methods would be more sophisticated, but this sort of reasoning is
useful for establishing back-of-the-envelope estimates of what is possible.}
In the latter case, we obtain 
\begin{equation}
\frac{\Delta C_\ell}{C_\ell}\approx 
\frac{1}{\ell }.
\end{equation}
We see that cosmic variance uncertainty is greatest at low-$\ell $ and rapidly
decreases with increasing $\ell .$ If we include detector noise and the resolution
of the survey due to beam smoothing, the above modifies to
\begin{equation}
\frac{\Delta C_\ell}{C_\ell}\approx 
\frac{1}{\ell } 
\left(
\frac{C_\ell +N_\ell }{ C_\ell }
\right) 
\end{equation}
where $N_\ell $ is the noise added by the detector and the quantum fluctuations of the 
incoming photon stream. The simplest noise model supposes white noise,
for which $N_\ell =N_0=\textrm{(constant)}.$ The simplest beam model 
supposes a Gaussian functional form, which attenuates the sky signal
by a factor of $\exp (-\sigma _\textrm{beam}^2\ell ^2/2),$ 
which has the same effect as assuming no attenuation and 
increasing the noise power spectrum by a factor of 
$\exp (+\sigma _\textrm{beam}^2\ell ^2),$ so that the above 
formula is modified to become  
\begin{equation}
\frac{\Delta C_\ell}{C_\ell }\approx 
\frac{1}{\ell } 
\left(
\frac{C_\ell +\bar N_\ell }{C_\ell }
\right) 
\label{CVeqnLast}
\end{equation}
where $\bar N_\ell =\exp (+\sigma _\textrm{beam}^2\ell ^2)N_0.$
This quantity increases with $\ell ,$ and when this 
quantity reaches order unity, this is the point beyond which
there is virtually no more statistically exploitable information.
The transition is rather abrupt because 
beyond $\ell \approx 500$ dampening effects combined with
the profile of the last scattering surface result in 
an exponential decay of the predicted $C_\ell $ spectrum. This 
effect can be mitigated to a certain degree with 
a very large telescope, so that the beam is very narrow, but 
beyond a certain point the primary CMB anisotropies 
(those emanating from the last scattering surface)
become subdominant and the so-called secondary 
anisotropies (e.g., from gravitational lensing, kSZ
from moving gas, and patchy reionization) take over.
All three effects share the same frequency 
dependence as the primary CMB anisotropies, so they
cannot be removed using standard foreground 
removal techniques. Other foregrounds such as 
compact sources also have spatial power spectra that dominate 
at large $\ell $ but these foregrounds can at least to some extent be 
removed by taking linear combinations of frequency
maps or more sophisticated methods of combining 
multifrequency data. 
 
The COBE DMR experiment
measured the CMB power spectrum relatively well in
the multipole range $\ell \lesssim 20 ,$ but at higher
$\ell $ the angular resolution was insufficient to
make a detection although 
extrapolations of the CMB angular power spectrum
that are very `blue' compared to a scale-invariant
spectrum could be excluded  
\cite{cobe-bandPower}.
The COBE instrument had an angular resolution of $7^\circ $
and the maps were smoothed to $10^\circ $ to guard against 
beam shape artefacts. 

The years following the COBE announcement were an exciting period because
the big question was how the CMB spectrum would look toward higher $\ell ,$
or equivalently on smaller angular scales. Broadly speaking, the COBE DMR results
provided a normalization factor for theoretical cosmological models,
and for some models the normalization found was problematic because
with the COBE normalization these models could not account
for the high value of $\sigma _8$ inferred from galaxy surveys at that
time.\footnote{$\sigma _8$ is defined as the standard deviation of the fractional mass fluctuation
today within a sphere of radius $8\,h^{-1}\textrm{Mpc}.$ Here
$h=H_0/(100\,\textrm{km}~\textrm{s}^{-1}\textrm{Mpc}^{-1})$ where $H_0$ is 
the value of the Hubble constant today.}
There was of course some wiggle room because of notions
such as `bias' reflecting that mass does not exactly trace light. But
a decisive test lay at higher $\ell ,$ where different models make 
starkly contrasting predictions.

\begin{figure}[t]
\begin{center} 
\includegraphics[width=12cm]
{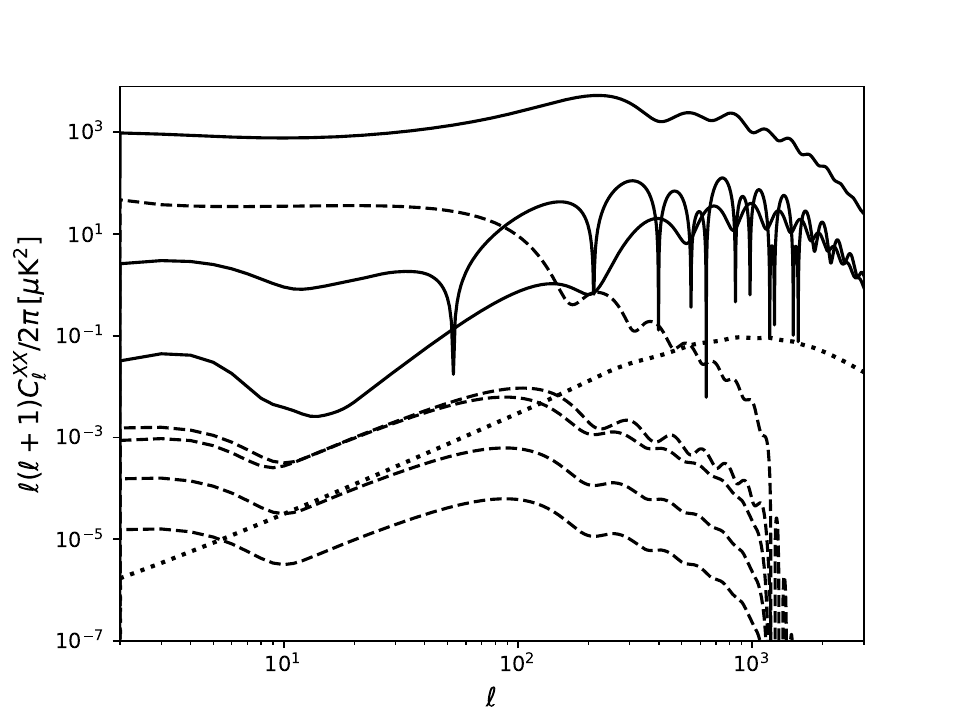}
\end{center}
\caption{{\bf Summary of T, E, and B Anisotropy Power Spectra from
Scalar and Tensor Modes.} The solid curves (from top to bottom)
represent the TT, TE, and EE CMB power spectra for the `scalar'
mode, while the dashed curves represent the CMB isotropies for the
`tensor' mode. The dashed curves (from top to bottom) and the
top solid blue curve represent the TT, TE, EE, and BB anisotropies
assuming a value of the tensor-to-scalar ratio of $r=0.1.$ The
lower two dashed curves represent the predicted BB anisotropy for
$r=0.01$ and $r=0.001.$ The dotted curve shows the BB power spectrum
arising nonlinearly from the `scalar' mode as a result of
gravitational lensing. 
[Credit: Reprinted with permission from Ref.~\citen{martinBook}. Courtesy of Cambridge University Press.]}
\label{AllSpectraPlot}
\end{figure}

\begin{figure}
\begin{center}
\includegraphics[width=0.45\textwidth]{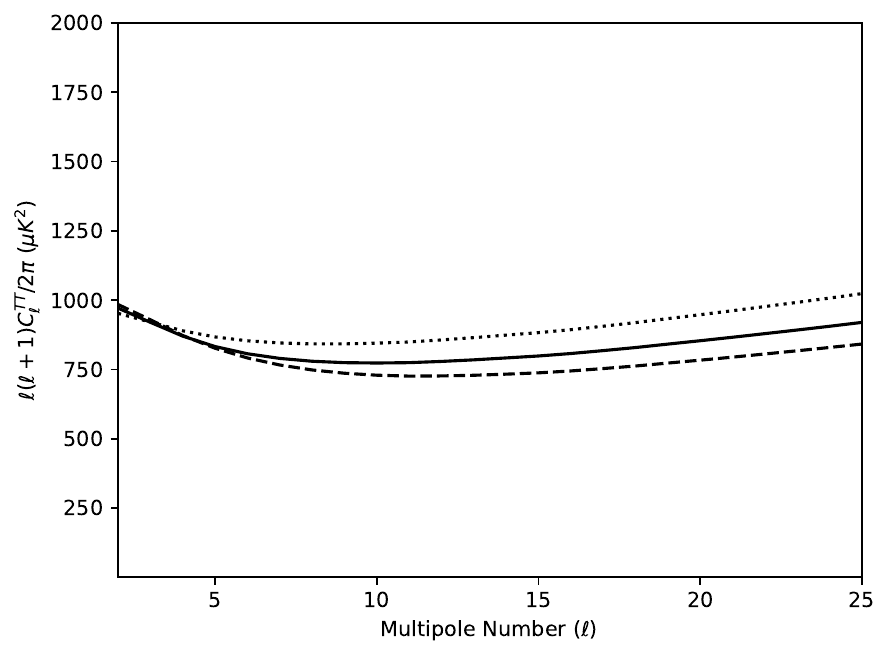}
\includegraphics[width=0.45\textwidth]{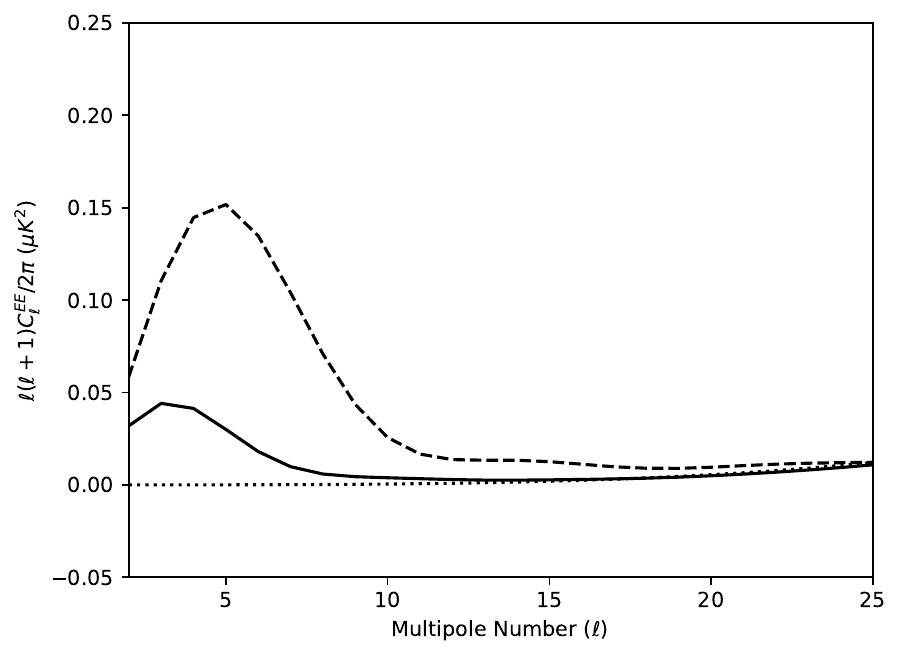}
\includegraphics[width=0.45\textwidth]{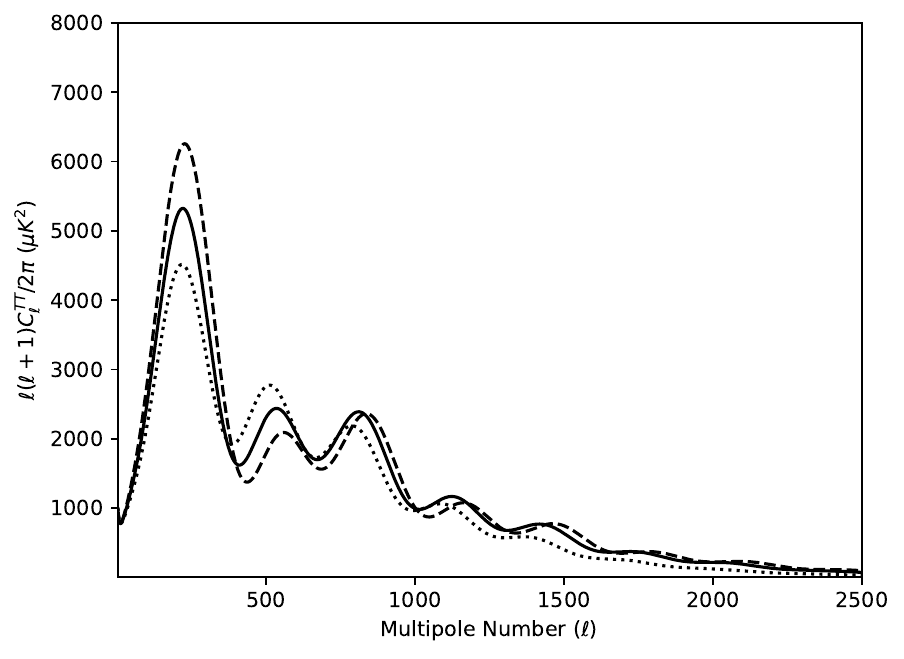}
\includegraphics[width=0.45\textwidth]{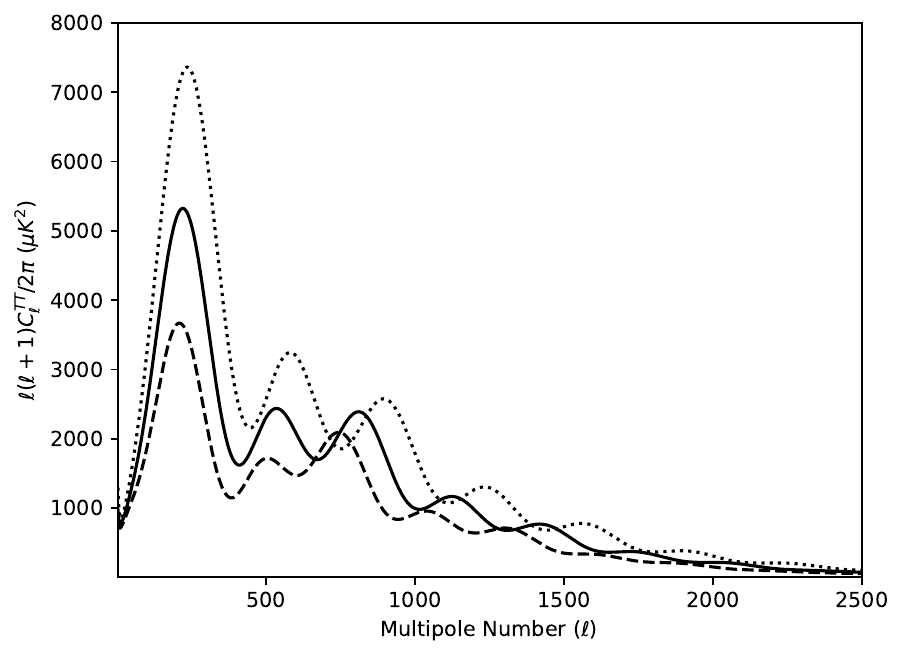}
\includegraphics[width=0.45\textwidth]{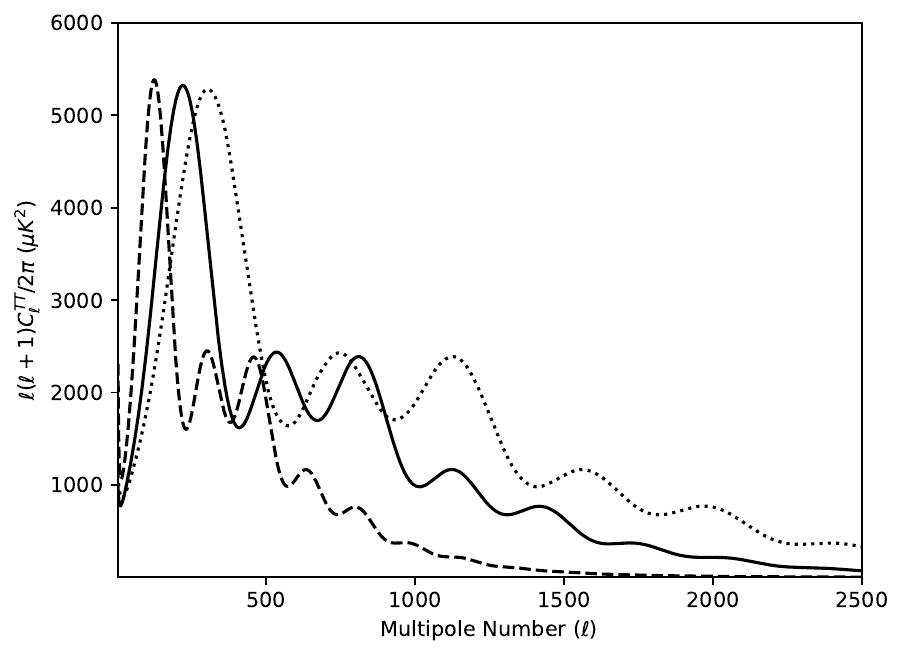}
\includegraphics[width=0.45\textwidth]{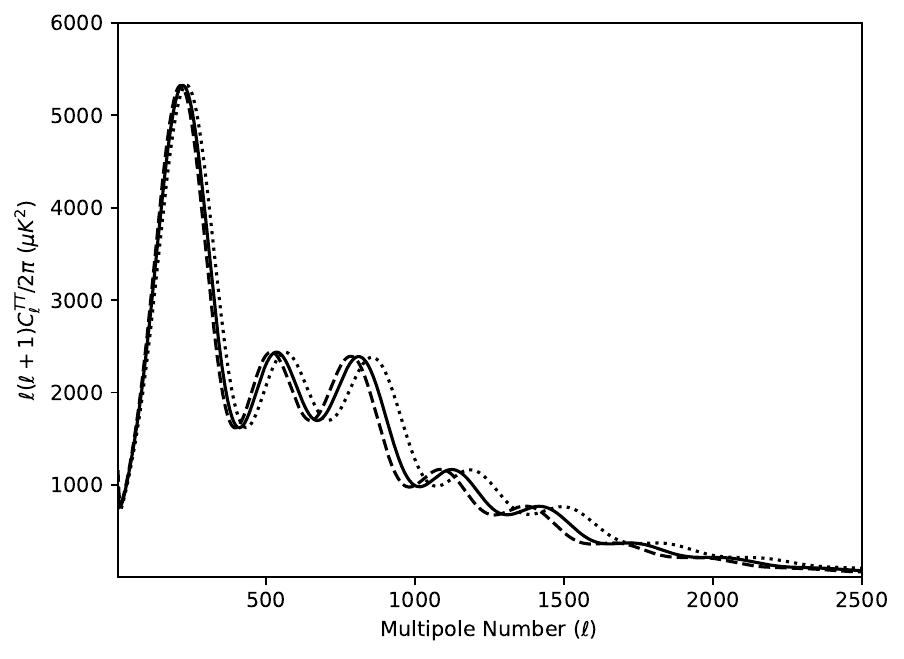}
\end{center}
\caption{%
\textbf{CMB Theoretical Power Spectra Shape Dependencies.}
The top two panels show the dependence of \smash{$C^{(S)\,TT}_\ell $} (left) and \smash{$C^{(S)\,EE}_\ell $} (right)
on the reionization optical depth $\tau = 0.0~\textrm{(dotted)},$
$0.06~\textrm{(solid)},$ and $0.12~ \textrm{(dashed)}.$ 
Except at very low multipoles $(\ell \lesssim 15)$ where there is hardly any attenuation,
\smash{$C^{(S)\,TT}_\ell $} is attenuated by a factor of $\exp (-2\tau ).$
For \smash{$C^{(S)\,EE}_\ell ,$} on the other hand, at very low $\ell $ a
reionization bump appears, corresponding
to rescattering by the reionized electrons.  In the middle row,
the dependences of \smash{$C^{(S)\,TT}_\ell $} on $\omega _b=\Omega _bh^2=$
0.011 (dotted), 0.022 (solid), and 0.033 (dashed) on the (left),
and on $\omega _c=\Omega _ch^2=$ 0.061 (dotted), 0.122 (solid), 0.244 (dashed)
on the (right) are indicated.  In the bottom row,
the dependences of \smash{$C^{(S)\,TT}_\ell $} on $\Omega _k=$ 
$+0.3$ (dotted), 0.0 (solid), $-0.3$ (dashed) on the (left), and on
$h=H_0/(100~\textrm{km}~s^{-1}~\textrm{Mpc}^{-1})=$ 0.50 (dotted), 0.67 (solid), 0.80 (dashed) on the (right) are shown.
[Credit: Reprinted with permission from Ref.~\citen{martinBook}. Courtesy of Cambridge University Press.]}
\label{Fig:PS-ShapeDependence}
\end{figure}

The $\Lambda $CDM model popular today combined with simple flavors of
inflation giving only adiabatic growing mode primordial cosmological
perturbations sufficiently well described by an approximately scale-invariant
power-law power spectrum predicts a series of peaks arising
from acoustic oscillations of the plasma with an exponentially decaying
damping tail. 
On the other hand, 
field ordering models, 
such as cosmic strings, global monopoles, or textures, 
do not predict such a series of sharp peaks. Rather both adiabatic 
growing and decaying modes excited continually throughout cosmic history
as a result of so-called `scaling behavior' rather than being excited
primordially at some time well before recombination, when the CMB perturbations
were imprinted. At that time theorists, who were largely inspired by considerations
of simplicity and aesthetics, favored a spatially flat universe with $\Omega _k=0.$
Nevertheless a variety of astronomical observations suggested that
$\Omega _m\approx 0.1 - 0.3,$ and this was before the observation
of Type Ia supernovae that seemed to rule out the theorists' favorite
$\Omega =1, \Lambda =0$ model. Since 
$\Omega _m+\Omega _k+\Omega _\Lambda =1$ at late times, 
$\Omega _k\approx 0.1 - 0.3$ was an option allowed by inflation under
the single bubble inflation models.\cite{gott,bgt,yst}
As indicated in Fig.~\ref{AllSpectraPlot}, these models predicted
a series of acoustic peaks, as predicted by the spatially flat inflation
models, but shifted to smaller angular scales because of the hyperbolic
spatial geometry, which acts like a sort of demagnifying lens. Moreover, 
the relative heights, positions, and detailed shapes of the acoustic
peaks are determined by a number of otherwise undetermined cosmological
parameters, as illustrated in Fig.~\ref{Fig:PS-ShapeDependence}. This was an exciting time, 
and the big question was which if any of these models would be able to fit the 
forthcoming data. 

In the years that followed COBE, a host of new data on smaller angular 
scales came in, at first with large error bars, so that compilations 
shown at conferences looked somewhat like the target of an archer barely
able to make the arrows strike the target. Theoretical speakers would
often cherry pick the observations that agreed with their models claiming
inside information regarding which points to believe.

\begin{figure}
\begin{center}
\includegraphics[width=0.47\textwidth]{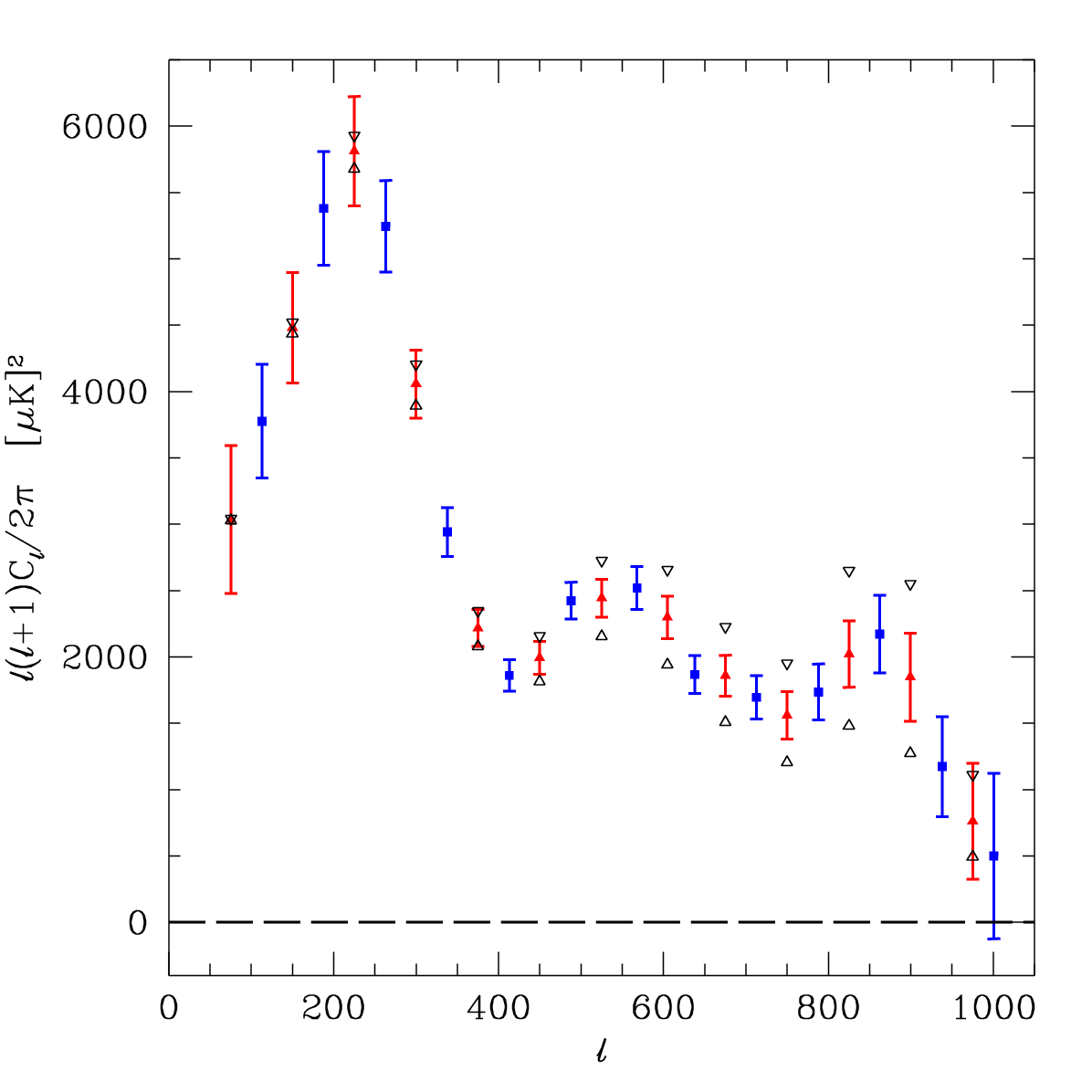}
\includegraphics[width=0.47\textwidth]{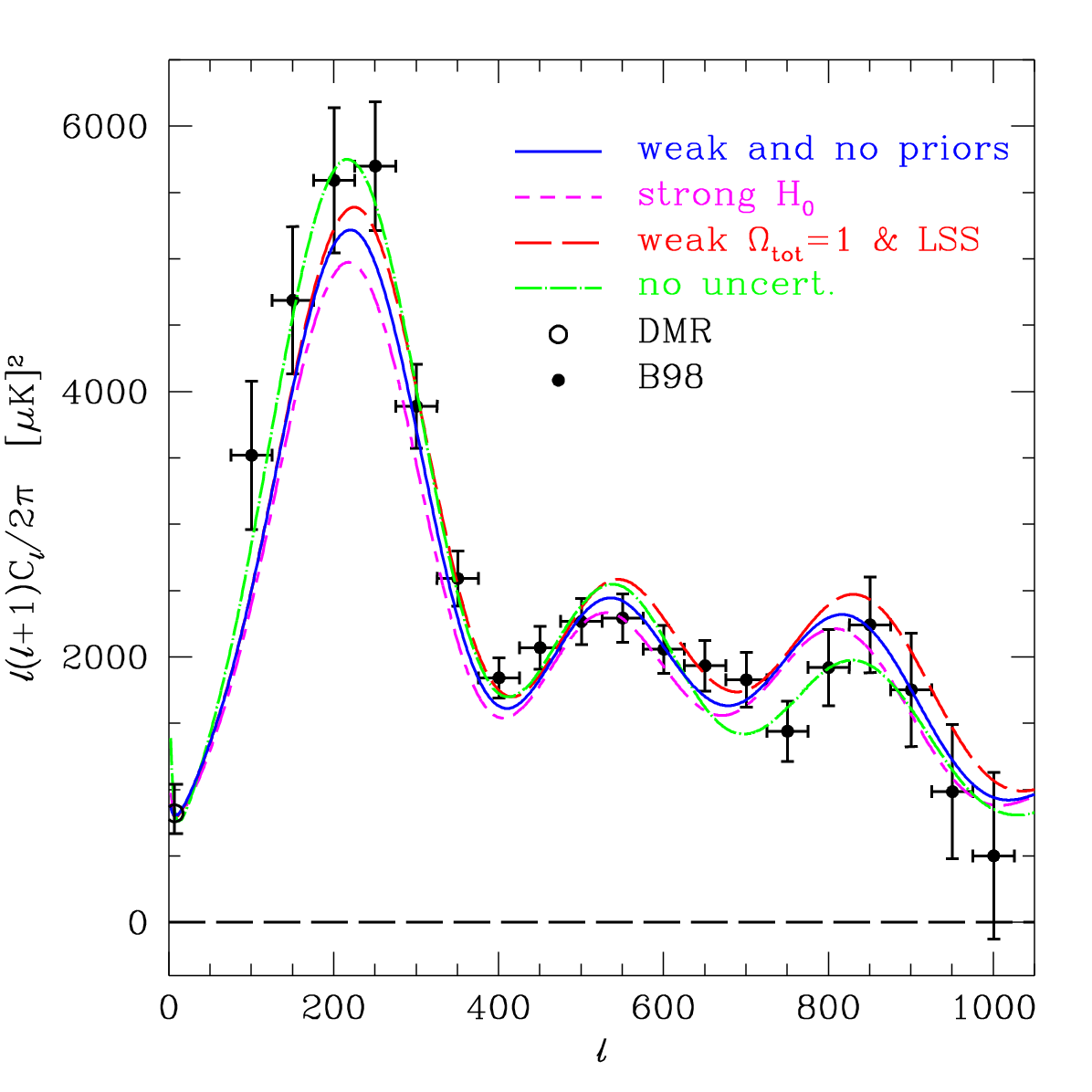}
\end{center} 
\caption{{\bf Boomerang Acoustic Peak Observation.} The $C^{TT}_\ell $
power spectrum as measured by Boomerang is shown on the left without a fit to a
theoretical model and on the right with the theoretical predictions for a
spatially flat cosmological model with an exactly scale invariant primordial power
spectrum for the adiabatic growing mode.
[{\it Credit: From Ref.~\citen{boom2Paper}. Courtesy of Boomerang Collaboration.}]
}
\label{BoomerangData}
\end{figure}

The convergence toward the observation of the first acoustic peak was 
incremental, but the observations from
Boomerang,\cite{boomPaper,boom2Paper}
Maxima,\cite{maximaRef} 
and TOCO\cite{tocoRef}
clearly indicated the presence of a first peak.
In Fig.~\ref{BoomerangData} 
we show the Boomerang team results, which  
together with other corroborating experiments
convinced most researchers that the first
Doppler peak lay as predicted by $\Omega _k=0$ inflation with an
approximately scale-invariant spectrum. 
This result ruled out defect models and models of inflation using a
hyperbolic geometry to keep $\Lambda =0,$ although models with
a subdominant contribution from topological defects to structure formation or models with
$\Omega _k$ today very near zero but not exactly zero could
technically speaking not be excluded.

One could argue that around this point precision observations of
the CMB began. There was a basic model that seemed to be able to
explain the observations. This was the six-parameter `concordance' 
model whose parameters are as follows:
$A_S,$ $n_s,$ $\Omega _b,$ $\Omega _m,$ $H_0,$ and $\tau .$
The question was whether this model would still be able to account for the observations
as further acoustic oscillations and the CMB damping tail
were mapped out.
Or would evidence of something new arise
proving that this is not the end of the story? Would the CMB 
determinations of the cosmological parameters agree with their determinations
using other means? 

\begin{figure}
\begin{center}
\includegraphics[width=0.7\textwidth]%
{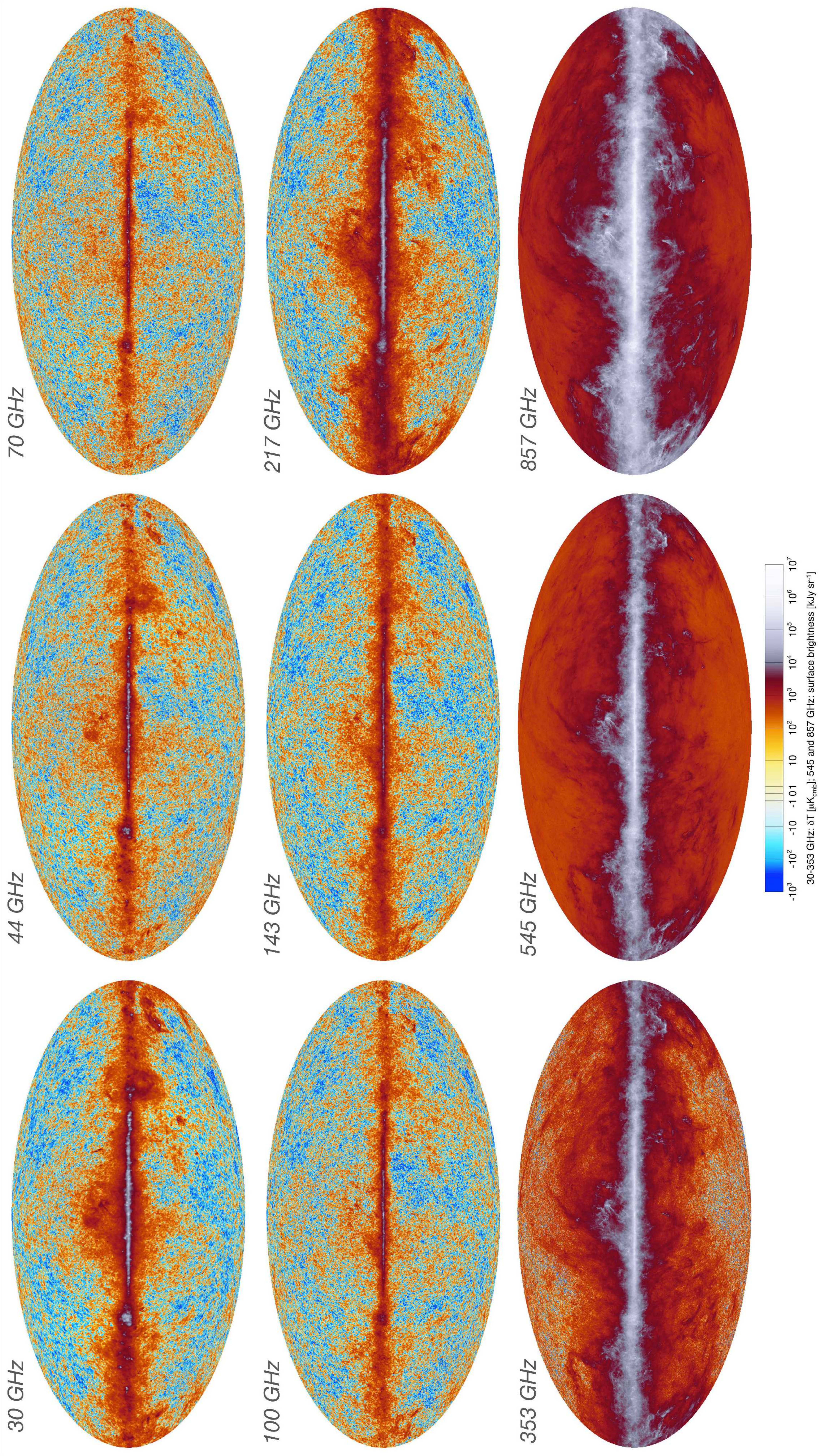}
\end{center}
\caption{%
\textbf{Planck Single-Frequency Temperature Maps.}
The single-frequency temperature maps are shown for the 9
Planck frequencies: 30, 44, 70, 100, 143, 217, 353, 545, and 
857 GHz. We observe that the Galactic plane is most visible 
in the lowest and highest frequency channels and least visible
in the 70 GHz channel. The maps are plotted in terms of CMB
thermodynamic temperature so that they would all look identical
if the CMB were the only component. In the low frequency maps,
the dominant foreground components are Galactic synchrotron emission
and free-free emission, whose Rayleigh--Jeans temperatures
scale as $\nu ^{-3}$ and $\nu ^{-2},$ respectively. At higher
frequencies thermal dust emission becomes the dominant 
contaminant. The thermal dust component may be approximated
$S(\nu )=\nu ^\alpha B(\nu ,T_\textrm{dust})$ where $T_\textrm{dust}\approx 20\,K,$
where $\alpha \approx 1-2.$ If the dust were optically thick, the 
$\nu ^\alpha $ factor would be replaced with unity. 
The strip about the Galactic plane is much narrower than in 
the COBE and WMAP images because of the higher angular resolution.
[Credit: ESA/Planck Collaboration, A\&A 641 (2020) A1. Reproduced with permission. \copyright ESO.]
}
\label{fig:PlanckTmp}
\end{figure}

\begin{figure}
\begin{center}
\includegraphics[width=0.99\textwidth]%
{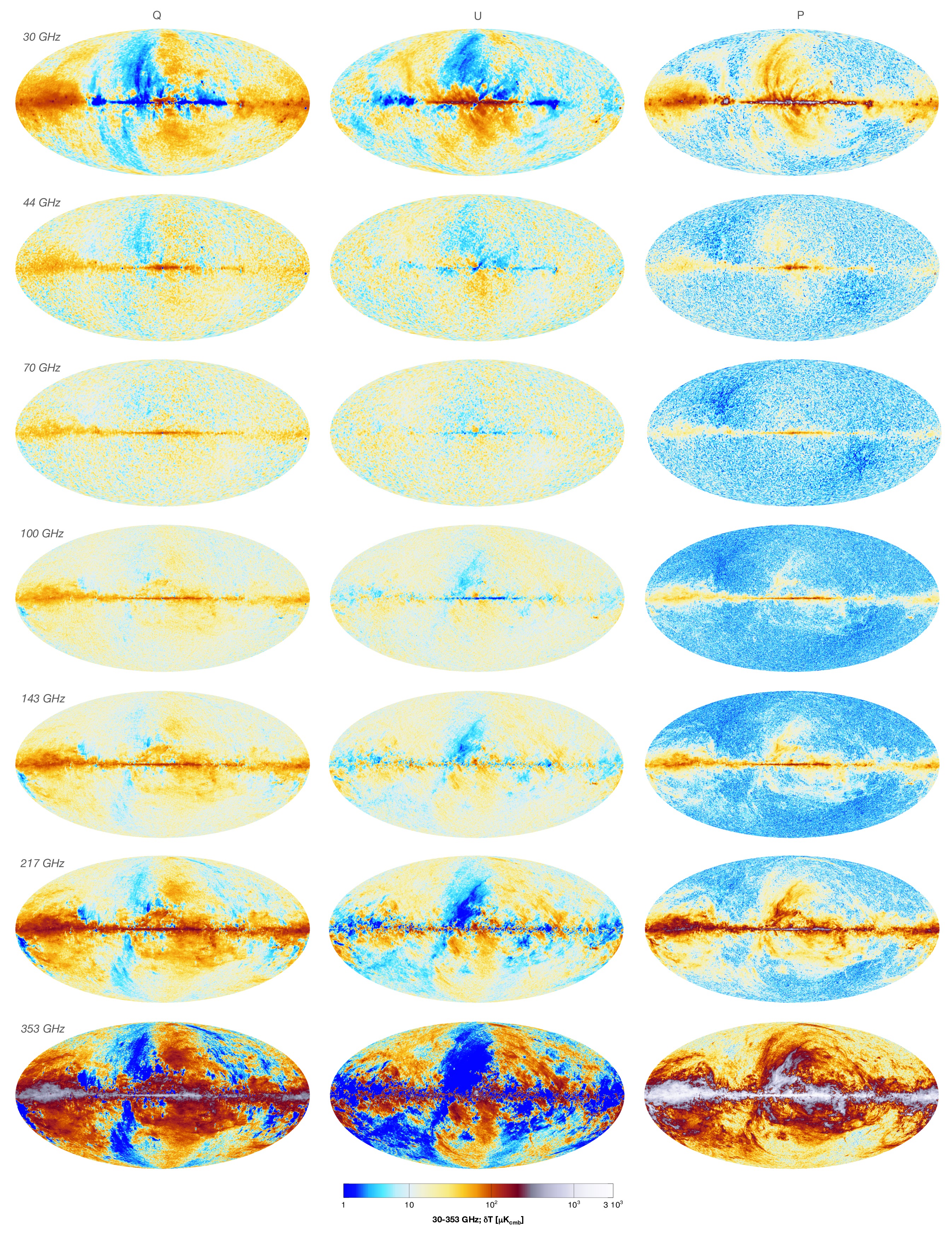}
\end{center}
\caption{%
\textbf{Planck Single-Frequency Polarization Maps.}
The linear polarization is described by two Stokes parameters $Q$ and $U,$ here defined
relative to the spherical basis vectors in Galactic coordinates having singularities
at the Galactic poles, and the total linear polarization is indicated by
$P=\sqrt{Q^2+U^2},$ which is basis independent. Each row corresponds to one of
the Planck frequency channels, with the $Q,$ $U,$ and $P$ sky maps shown (left to right).
In these maps the Galaxy is more visible and thus extends farther from the equator
because the polarization fraction of the contaminants is larger than that of the 
primordial CMB signal. 
[Credit: ESA/Planck Collaboration, A\&A 641 (2020) A1. Reproduced with permission. \copyright ESO.]}
\label{fig:PlanckPol}
\end{figure}

\begin{figure}
\begin{center}
\includegraphics[width=0.7\textwidth]%
{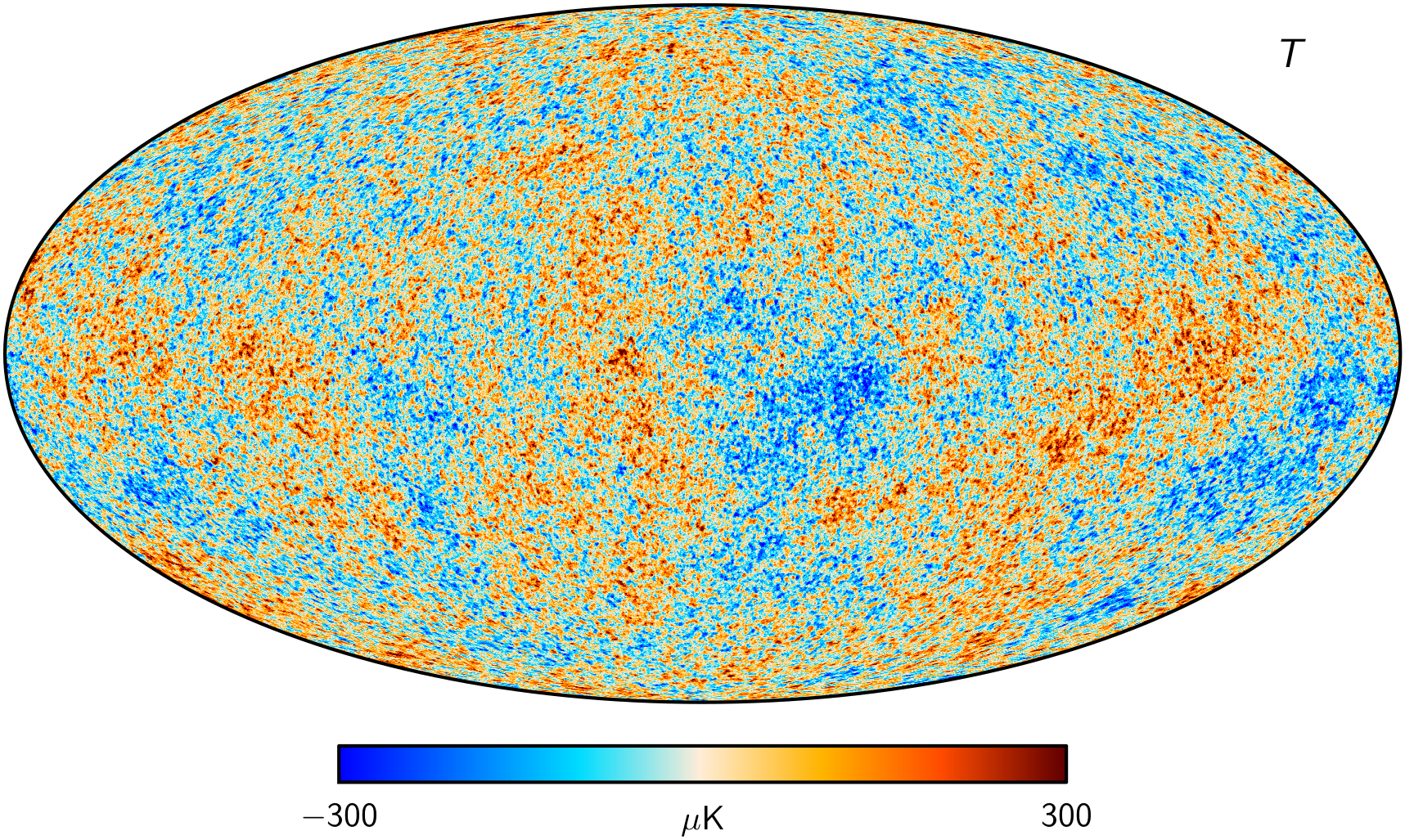}
\end{center}
\caption{%
\textbf{Planck CMB-Only Map.}
Since each of the components contributing to the single-frequency microwave sky maps
has a different frequency dependence that can roughly be factorized as product of
the form $S(\nu )M(\hat {\boldsymbol{\Omega }}),$ a nearly pure CMB-only map may be
constructed by taking appropriately weighted linear combination of single-frequency
maps. The result is seen above and we observe that the Galactic plane, visible to
some extent in all the single-frequency maps, is no longer apparent to the unaided 
eye, and the CMB now looks truly looks isotropic.
[Credit: ESA/Planck Collaboration, A\&A 641 (2020) A1. Reproduced with permission. \copyright ESO.]
}
\label{fig:PlanckCMB-Only}
\end{figure}

\begin{figure}
\begin{center}
\includegraphics[width=0.9\textwidth]%
{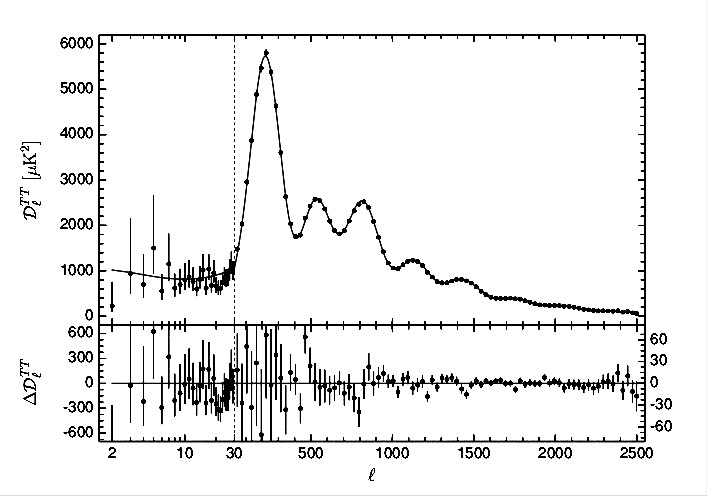}
\end{center}
\caption{%
\textbf{Planck 2018 Temperature Power Spectrum.}
Plotted is $\mathcal{D}_\ell ^{TT}=\ell (\ell +1)C_\ell ^{TT}/(2\pi )$ which 
can be understood as the square power per logarithmic wavenumber (i.e.,
$d(T^2)/d[\log \ell ]$) so that a scale-invariant spectrum would
be represented by a horizontal line. The bottom plot shows the residuals
with respect to the best-fit theoretical model. On large angular scales,
cosmic variance dominates, becoming smaller on intermediate scales until
the measurements become limited by a combination of detector noise and
insufficient angular resolution. The fact that no pattern can be seen
in the residuals suggests that the theoretical model offers an
adequate explanation of the data.
[Credit: ESA/Planck Collaboration, A\&A 641 (2020) A6. Reproduced with permission. \copyright ESO.]
}
\label{planckTT}
\end{figure}

\begin{figure}
\begin{center}
\includegraphics[width=0.7\textwidth]%
{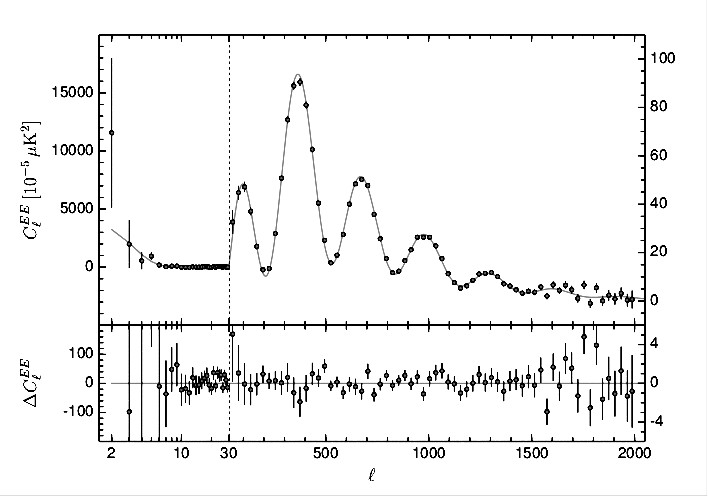}
\includegraphics[width=0.7\textwidth]%
{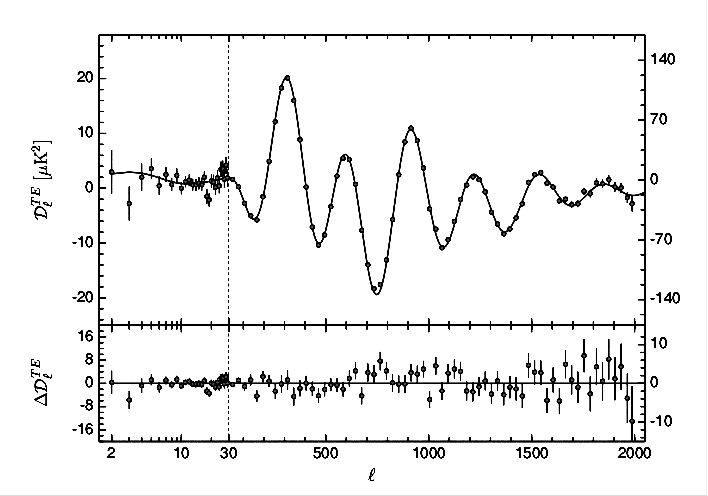}
\end{center}
\caption{%
\textbf{Planck 2018 EE and TE Power Spectra.}
The top panel shows the $C_\ell ^{EE}$ [note the expanded scale 
for $\ell <30$]. At very low $\ell $ (i.e., $\lesssim 7$) we see 
the `reionization bump,' which allows us to estimate the reionization
optical depth $\tau .$ The sequence of acoustic peaks at higher $\ell $
and their shapes and amplitudes allow corroboration of the model and 
cosmological parameters from the temperature-only power spectrum.
Below is the $TE$ cross-correlation power spectrum. 
[Credit: ESA/Planck Collaboration, A\&A 641 (2020) A6. Reproduced with permission. \copyright ESO.]
}
\label{planckTEandEE}
\end{figure}

WMAP\cite{wmapDescr}
 was a US NASA follow up to COBE designed to map the entire 
microwave sky at five frequencies at both greater sensitivity and
greater angular resolution.\footnote{%
A full bibliography of the official WMAP Collaboration papers may be found on the webpage
https://lambda.gsfc.nasa.gov/product/wmap/dr5/map\_bibliography.html}
Like COBE, WMAP used HEMT amplifiers, 
but rather than pointing horns directly at the sky as in COBE, WMAP
used two back-to-back telescopes 
(with a 1.4\,m $\times $ 1.6\,m primary mirror and an off-axis Gregorian configuration)
allowing for an enhanced angular 
resolution. Like COBE, WMAP rapidly switched between directions
in the sky separated by $\approx 100^\circ ,$ using only difference measurements 
to construct microwave sky maps at each frequency. WMAP observed
in five frequency channels centered at 
$23,$
$33,$ $41,$ $61,$ and $94$ GHz
with beam widths (fwhm) 
of $53,$ $40,$ $31,$ $21,$ and $13$
arcmin, respectively. Unlike COBE WMAP was situated at L2
(the second Lagrange point, situated far from both the earth and the moon) 
rather in a low-earth orbit, thus avoiding interference from
the earth and the moon and providing exquisitely stable
observing conditions. 

WMAP extended the observations of the acoustic oscillation 
toward larger $\ell ,$ and moreover provided full-sky polarization
maps allowing the $TE$ and $EE$ power spectra to be characterized. 
The WMAP results also allowed a determination of the reionization
optical depth, the non-Gaussianity parameter $f_{NL},$ and
provided the first evidence for a small deviation from exact scale invariance
(i.e., $n_S\ne 1$).\cite{wmapThreeYearCParams}
Moreover, the polarization data subsequently reported by WMAP in 2006
confirmed the same basic story as suggested by the temperature anisotropies.\cite{wmapThreeYearPolar,wmapThreeYearCParams}

At the WMAP 2003 press conference John Bahcall \cite{bahcall}
aptly stated, 
``the most revolutionary result was that there were no revolutionary results.'' 
Jumping ahead a bit, this same characterization also applies to the Planck Results.
While the data from Boomerang and other experiments unequivocally showed the existence of the 
acoustic peak with hints of subsequent peaks at higher multipole number,
WMAP and Planck mainly confirmed the basic model suggested by Boomerang,
Maxima, and TOCO albeit at much greater precision. 
To some with a more revolutionary spirit, this outcome was a disappointment, 
and 
upon closer scrutiny 
a failure 
of the simple model that seemed to
account for the observations would have been the preferred outcome, 
with theorists sent back to the drawing board.
In the course of the 20th century cosmology there have been a number of unexpected 
results (e.g., expansion of the universe, dark matter, dark energy) forcing
a major revision of our understanding of the Universe. 
On the other hand, for others there is great satisfaction in confirming a model 
at ever greater precision, much as has happened with QCD and the standard 
electroweak model, for which the discovery of the Higgs particle
and a minimal Higgs sector was simultaneously a great triumph
and a disappointment that low-energy SUSY or some other extension of the standard 
model was not confirmed.

The ESA Planck mission \cite{planck2013overview,planckOverview2018}
consisted of two instruments: a Low Frequency Instrument (LFI),
consisting of three frequency channels (centered at 30, 44, and 70 GHz)
and using a HEMT technology similar to that used by WMAP; and a 
High Frequency Instrument (HFI), using bolometric detection technology
with six bands at 
100, 143, 217, 353, 545, and 857 GHz.\footnote{%
A complete list of the Planck Collaboration publications may be found on the webpage 
https://www.cosmos.esa.int/web/planck/publications}
All channels except for the two highest were polarization sensitive.
Figures 
\ref{fig:PlanckTmp} 
and 
\ref{fig:PlanckPol} 
show the Planck temperature and 
polarization maps, respectively, for each of these channels.
Figure \ref{fig:PlanckCMB-Only} shows a CMB-only maps obtained by combining the 
single-frequency maps in such a way as to block the foreground emissions allowing
only the primordial CMB signal to pass. 

Bolometric technology has the drawback of requiring active cooling. The
Planck HFI bolometers operated at 100 mK, whereas WMAP was passively cooled
to $\approx 90$\,K. Bolometers however are less noisy than coherent detectors and can 
operate above 100 GHz (which is approximately the point beyond which coherent
detection is no longer feasible). Observations at higher frequency allow for higher 
angular resolution given
the same mirror size and also allow characterization of the dust Galactic foreground 
emission, for which WMAP had to use other external templates at much 
higher frequencies leading to a larger extrapolation error.

Figures \ref{planckTT} and \ref{planckTEandEE} show the power spectra from the Planck 2018
release.\cite{planckOverview2018}
We see much the same patterns as those seen by WMAP but with smaller error bars
and extending out to higher $\ell .$
Planck measured the CMB temperature power spectrum up to approximately
$\ell \!=\!2500$ identifying the precise positions and amplitudes of
seven acoustic peaks and seven acoustic troughs of the angular
power spectrum. The troughs and peaks of the
$C_\ell ^{EE}$ and
$C_\ell ^{TE}$
power spectra 
were also mapped out with high precision although not to as
large $\ell $ owing to the lower amplitude of the polarized signal. 
And importantly, the
$C_\ell ^{EE}$ and
$C_\ell ^{TE}$
power spectra when analyzed without the
$C_\ell ^{TT}$
power spectrum
gave cosmological parameter determinations consistent with those
obtained using $C_\ell ^{TT}$ alone.

\begin{table}
\tbl{\textbf{Six-Parameter Baseline Cosmological Model.}
The central values and $1\sigma $ errors for the concordance model parameters based on the 
Planck 2018 TT, TE, EE + low-E +
lensing likelihood are given.
[Adapted from Table 1 in Ref.~\citen{2020A&A...641A...6P}.]
}
{\begin{tabular}{|l|l|}
\hline
\hline
Parameter & Value \\
\hline
\multicolumn{2}{c}{Six Baseline Parameters}\\
\hline
$\Omega _b\,h^2$   &
$0.02233 \pm 0.00015$ \\
$\Omega _c\,h^2$    &
$0.1198 \pm 0.0012$  \\
$10^2\,\theta _{MC}$ &
$1.04089 \pm 0.00031$\\
$\tau $ &
$0.0540 \pm  0.0074$ \\
$\ln \left( 10^{10} \mathscr{A}_S \right)$ &
$3.043 \pm 0.014$ \\
$n_s$     &
$0.9652 \pm 0.0042$ \\
\hline
\multicolumn{2}{c}{Derived Parameters}\\
\hline
$\Omega _mh^2$ &
$0.1428 \pm 0.0011$ \\
$H_0\,(\textrm{km}\,s^{-1}\,\textrm{Mpc}^{-1})$ &
$67.37 \pm 0.54 $ \\
$\Omega _m$ &
$0.3147 \pm 0.0074$ \\
$t_0\, (10^9 \textrm{yr})$ &
$13.801 \pm 0.024$ \\
$\sigma _8$ &
$0.8101 \pm 0.0061$ \\
$S_8= \sigma _8(\Omega _m/0.3)^{1/2}$ &
$0.830 \pm 0.013$ \\
$z_\textrm{rec}$  & 
$7.64 \pm 0.74$ \\
$100~\theta _*$ &
$1.04108 \pm 0.00031$ \\
$r_\textrm{drag}~ (\textrm{Mpc}) $ &
$147.18 \pm 0.29$ \\ 
\hline
\end{tabular}}
\label{Table:P2018Params}
\end{table}

The full-sky maps obtained also made possible analyses beyond the
power spectrum. Constraints on the amount of bispectral anisotropy
(which vanishes in a Gaussian cosmological model) were obtained for
various theoretical templates, and the hint of a nonzero bispectral
signal from WMAP\cite{wmapNineYear}
$f_{NL}^{\textrm{local}}=37.2\pm 19.9$
for the `local' template was not confirmed. Instead
a significantly lower upper limit consistent with zero 
$f_{NL}^{\textrm{local}}= 2.7\pm 5.8$
was obtained.\cite{planckNG2013,planckNG2018}
Various tests of statistical isotropy were applied to the CMB-only
maps and signs of dipolar modulation pointed out by WMAP (and also
COBE) were confirmed at modest statistical significance. However
such modulation seems to affect only the low multipoles (i.e.,
$\ell \lesssim 100$) and no 
statistically significant modulation was found at higher
multipoles.\cite{planckIsoStat2013,planckIsoStat2018}

The Planck data is broadly consistent with the above mentioned
six-parameter concordance model, for which the fitted parameter
values and their errors are given in Table \ref{Table:P2018Params}. 
\cite{planck2013params,2020A&A...641A...6P}
Notably the determination of the scalar primordial power spectrum
spectral index 
$n_S= 0.9649\pm 0.0042$ rules out at about $8\sigma $ an exactly scale invariant
primordial power spectrum. Most inflationary models favor a slightly
red spectrum (compared to a scale invariant spectrum), although inflationary
models can be constructed giving a blue spectrum.
The Planck data also placed constraints on specific inflationary models.
\cite{planckInflation2013,planckInflation2018}
These constraints can be expected to tighten as better upper bounds
on $r$---or possibly a first detection---are obtained by future experiments.
Constraints on various extensions of this base model were also obtained.
\cite{planck2013params,2020A&A...641A...6P}
In particular, the number of relativistic degrees of freedom (expressed
as a number of light neutrino species) was constrained to 
$N_\textrm{eff}=2.99\pm 0.17,$ and the neutrino mass was constrained 
to satisfy $\sum m_\nu < 0.12\, \textrm{eV}.$

Using the small-angle structure in the CMB sky maps, Planck also reconstructed the
gravitational lensing potential and from the reconstructed lensing 
maps measured the gravitational lensing power spectrum, resulting
in a detection of gravitational lensing of the CMB 
at $40\,\sigma $ in the temperature maps and at $5-9\,\sigma $ in the polarization
maps.\cite{planckLensing2018} 
CMB lensing was first detected using the cross-correlation between a noisy map
of the lensing inferred from a CMB map  and a map of the LSS projected onto
the celestial sphere.\cite{smithLensing}
Gravitational lensing data is useful
in estimating the magnitude of the perturbations in the matter power spectrum
and also constraining its shape. As discussed further below, data of
the gravitational lensing of the CMB should greatly improve in the future,
as more modes on smaller angular scales, as measured from the ground, will
significantly tighten error bars. Lensing reconstruction using the $B$ mode polarization
suffers much less from cosmic variance than lensing reconstruction using 
temperature maps. Gravitational lensing with better data also promises to provide
constraints on absolute neutrino masses. 

A puzzle remains as to how to 
understand the discrepant measurements of the present value of the Hubble
constant $H_0.$  Determinations of $H_0$ made using the CMB anisotropies---in other words, 
probes of the early universe---give a lower value
than determinations of $H_0$ made in the more recent universe by establishing 
a cosmic distance ladder. This discrepancy is often referred to as the `$H_0$ tension.'
The Planck 2018 cosmological parameters paper for example reports
$H_0=67.4\pm 0.5\, \textrm{km}~\textrm{s} ^{-1}\,\textrm{Mpc} ^{-1}$
whereas Riess et al.\cite{riessShoes}
using a distance ladder constructed using parallax, Cepheids, and Type Ia supernovae
find 
$H_0=73.30\pm 1.04\, \textrm{km}~\textrm{s} ^{-1}\,\textrm{Mpc} ^{-1}.$
These results are discrepant by over $5\,\sigma .$ Other CMB results
give similar low values of $H_0,$ and other cosmic distance ladder determinations
give similarly larger values although in some cases lower than in Riess et al. For an detailed discussion, see 
the Di Valentino et al. reviews.\cite{valentinoA,valentinoB}
It is not clear whether this tension is a sign that the present cosmological
model in not the full story, or whether the explanation rather lies
in unknown systematic errors or an inproperly estimated error budget.

\section{Ground-Based and Balloon-Based Observation}

While observations of the CMB from space, as carried out by
COBE, WMAP, and Planck,
benefit from the absence of atmospheric interference and
remarkably stable observing conditions over long periods
of time, observations from the ground can be carried out with
much larger aperture telescopes. Because CMB observations are
diffraction limited, this allows for vastly superior angular
resolution. Consequently the most stringent constraints on
cosmological models are obtained by combining observations
from space (for example from Planck) for the large angular
scales (or equivalently, low $\ell $) with observations from
the ground for the small angular scales (or high $\ell $).

Following COBE there were numerous ground- and balloon-based experiments seeking to measure
the predicted acoustic peaks on smaller angular scales. Boomerang, MAXIMA, and TOCO were mentioned
above, but there were many other experiments too numerous to list here, and  
the NASA LAMBDA website provides a rather complete list of these
CMB experiments,\footnote{https://lambda.gsfc.nasa.gov/product/expt/}
which beyond their measurements of the CMB anisotropies played an important
role in technology development, as it took some time to perfect the instrumentation
used in space and in more modern ground-based experiments.

The best more recent observations of the CMB from the ground are
carried out either from the Atacama desert in Chile or from
the South Pole. At both sites the amount of precipitable
water vapor is particularly low compared to other sites, leading
to better atmospheric transparency and lower parasitic emission
from the atmosphere.

The Atacama Cosmology Telescope (ACT) used a 6.0 m aperture Gregorian telescope 
to survey the sky at microwave wavelengths. The telescope was installed in 2007
at Cerro Toco in the Atacama desert in Chile and 
upgraded to ACTPol (2013–2016) and then to Advanced ACT (2017–2022)
until its decommissioning in 2022. 
The most recent cosmology results from ACT may be found in Refs.~\citen{actParams2025} and \citen{act-calabrese}. 
Using determinations of the power spectrum up to $\ell \approx 4000,$ far beyond the angular resolution
of Planck, allowed additional acoustic peaks and the structure of the damping tail to be explored.
Other ACT science includes measurement of the gravitational lensing power spectrum,\cite{actQuSherwin}
establishing a catalog of galaxy clusters detected by means of the thermal Sunyaev--Zeldovich (tSZ) effect,\cite{actTSZ}
and the detection of the kinetic Sunyaev--Zeldovich (kSZ) effect.\cite{actKSZ} 
The Sunyaev--Zeldovich effects arise from hot ionized gas in galaxy clusters. The electrons of this fully ionized
gas rescatter the CMB photons. The thermal Sunyaev--Zeldovich arises from the high temperature of the gas, which 
on the average acts to inject energy into the CMB photons through a random Doppler shift. At low frequencies
(for $\nu \lesssim 217$ GHz), somewhat paradoxically this heating lowers the brightness temperature of the rescattered CMB photons
because low frequency photons are depleted, whereas at high frequencies ($\nu \gtrsim 217$ GHz) the brightness temperature
of the CMB is raised. The kinetic Sunyaev--Zeldovich (kSZ) effect arises from the peculiar velocity 
of the cluster gas (with respect to the CMB rest frame), which likewise changes the photon frequencies through
the Doppler effect. The kSZ effect is very difficult to detect for two reasons:
(1) The change in brightness temperature is smaller than the kSZ, and (2) 
the spectrum of the change in temperature has the same frequency dependence as the primary CMB anisotropies,
whereas the tSZ has its own spectral shape, distinct from that of the primary CMB anisotropies and of the other foreground
emissions processes.
 
The Huan Tran Telescope (HTT), a 3.5 m Gregorian telescope, was part of a project to measure the polarization 
of the cosmic microwave background radiation.  
Attached to the telescope was the POLARBEAR experiment, which was an 
array of cooled bolometers. The HTT was first installed 
in 2010 for testing at the CARMA site and was moved to a location on Cerro Toco near the ACT in 2011. The HTT,
developed by a consortium led by the University of California, Berkeley, 
saw first light in January 2012. In 2014 POLARBEAR observed $B$ modes due to gravitational lensing at 1.8 $\sigma.$\cite{polarbearGL}
POLARBEAR seeks to map out the $B$ modes, both of possible primordial origin at lower $\ell $ and those
at higher $\ell $ due to gravitational lensing by matter inhomogeneities.\cite{polarbearGL2}

The South Pole Telescope (SPT) is a 10 m telescope sited at the Amundsen–Scott South Pole Station 
in Antarctica. SPT achieved first light in 2007 and since that time has produced a stream of
interesting results, similar to those of ACT including CMB temperature and polarization power
spectra at small angular scales and their implications for cosmology, \cite{spt1,spt2}
detection using cross-correlation of $B$ modes from gravitational lensing of $E$ modes,\cite{sptpol1}
and Sunyaev--Zeldovich catalogs and other SZ science.\cite{sptSZ} 

In China a new experiment AliCPT \cite{aliCPT,aliCPT-mainPaper2,aliCPT-mainPaper}
is in the process of being deployed in Tibet, although no results papers have yet appeared.

The more recent Simons Observatory\cite{simonsObs-SciGoalsAndForecasts,spt3G}
is a microwave observatory using a 6\,m Large Aperture Telescope (LAT) 
and three 0.5 m Small Aperture Telescopes (SATs) with a total of 60,000 transition-edge sensor 
bolometers. These telescopes will conduct a wide survey of the cosmic microwave background and its linear polarization 
with arcminute resolution and a sensitivity an order of magnitude higher than the Planck space telescope. 
The SATs began observing in April 2024, while the LAT's started taking data in early 2025.
In parallel, experimentation at the South Pole continues. 

\section{CMB Polarization}

The simplest theoretical calculation 
of the predicted CMB anisotropy 
uses the Sachs--Wolfe approximation
\cite{sachs1967perturbations}
under which the universe transitions instantaneously from a completely
opaque state  (due to frequent Thomson scattering of photons off free electrons) 
to a transparent state (because neutral atoms hardly scatter low
energy photons). Under this approximation, the visibility function 
is a Dirac $\delta $-function. The CMB [except for an Integrated Sachs--Wolfe (ISW)
contribution] offers a view of the state of the universe on an infinitely
thin spherical shell known as the `last scattering surface' at $z\approx 1100.$ 

But this is not the whole story. Recombination and decoupling of the CMB
proceed quickly but not instantaneously. Therefore the last scattering surface
has a nonzero thickness. Because Thomson scattering is polarization dependent,
the CMB observed by us today is partially polarized. There is no circular
polarization, only a partial linear polarization, described by the Stokes
parameters $Q$ and $U.$ Roughly speaking, the polarization seen today in a given 
direction on the sky corresponds to the quadrupole temperature anisotropy
seen by the electrons
at last scattering. Quantitatively, this is a bit too simple,  
because this radiation 
incident on the electrons of last scattering 
is also partially polarized. A precise
calculation requires tracking multiple scatterings through solving
a Boltzmann hierarchy numerically.\cite{peeblesYu,cpMa} The scale at which $C_\ell ^{EE}$ 
peaks corresponds roughly to the angular scale of 
the distance between last and next-to-last scattering. On very small scales
the averaging over radial positions of the electrons of last scattering
acts to wash out the polarized anisotropy. 
On larger scales there is also suppression 
because the photons do not travel far enough
to develop a large enough quadrupole anisotropy.

The polarization probes details of recombination not accessible
by observing the temperature anisotropy alone. Polarization observations
allow degeneracies to be broken (for example in testing the hypothesis 
of adiabaticity of the primordial perturbation) and moreover allow
corroboration of the model inferred from $C_\ell ^{TT}$ alone.

Another feature of CMB polarization is that it allows us to 
probe reionization and consequently fix the overall amplitude
of the primordial perturbations. 
It is now known that the universe today has been almost
completely reionized (except in certain regions
of especially high baryon density). Ionizing radiation from the first generation of stars and quasars
and from subsequent star formation suffices to maintain almost all the hydrogen in an ionized 
state. From the Planck data we know that reionization occurred around $z_\textrm{rec}=7.5$ although
CMB data alone is insufficient to map out the detailed history of how reionization occurred.
On small angular scales reionization has the effect of erasing the CMB anisotropies. The
reionization optical depth according to the Planck 2018 results is $\tau =0.054\pm 0.007,$
meaning that approximately one out of every 20 photons emanating from the surface of last
scatter is rescattered by a free electron from reionization. On small scales the CMB
anisotropies are erased due to a smearing effect as the scattering by free electrons
from reionization 
has the effect of averaging 
over a 
large region. Only on the largest angular scales do the temperature anisotropies persist.
The situation resembles looking at something through a piece of frosted glass. Silhouettes
remain visible while details are erased. This is what happens for the temperature anisotropy
on small scales.

For the CMB polarization, reionization produces an enhancement on large scales known as
the `reionization bump.' On scales larger than $x_\textrm{ls-nls},$ the order of magnitude
of the comoving distance between last and next-to-last scattering,  
the polarization is proportional to \smash{$(x_\textrm{ls-nls})^2(\partial ^2T/\partial x^2).$}
The polarization on large scales arising during the epoch of recombination is small
because 
\smash{$x_\textrm{ls-nls}^\textrm{(rec)}$} is small---it is of order the width of the last scattering
surface. But 
\smash{$\tau (x_\textrm{ls-nls}^\textrm{(reion)})^2(\partial ^2T/\partial x^2)$}
is large despite the smallness of the $\tau $
because 
${x_\textrm{ls-nls}^\textrm{(reion)}}$
is huge: it is a good fraction of the size of current horizon. Observing 
the $E$ mode polarization at very low $\ell $ allows a relatively accurate determination of
the reionization optical depth, in principle limited only by cosmic variance.

The first experiment to detect the polarization of the CMB was the DASI \cite{dasiPaper}
interferometric experiment located in the Owens Valley in 
California, USA. WMAP and Planck subsequently measured the 
CMB polarization with greater precision on large and intermediate
angular scales, 
and ground-based experiments are mapping the polarization on smaller angular scales. 

\section{$E$ Modes and $B$ Modes}

The search for a $B$ mode of the 
polarization anisotropy of primordial origin is one of the main 
areas of intense activity of contemporary CMB observations for several
reasons that we now explain.

The polarization of the microwave sky may be characterized using 
the $Q$ and $U$ Stokes parameters. This
parameterization however involves
an arbitrary choice of basis and thus is not useful for making contact between 
theory and competing cosmological models. More useful
is a decomposition into $E$ modes and $B$ modes, which may defined
as follows using the scalar spherical harmonics as a starting point:
\begin{equation}
\begin{aligned}
E_{\ell m,\,ab}(\hat {\boldsymbol{\Omega }}) 
&=N_\ell ^E
\left(
\nabla _a
\nabla _b
-\frac{1}{2}
\delta _{ab}
\nabla ^2
\right) Y_{\ell m}
(\hat {\boldsymbol{\Omega }}),\\
B_{\ell m,\,ab}(\hat {\boldsymbol{\Omega }}) 
&=N_\ell ^B 
\frac{1}{\sqrt{2}} 
\left(
\nabla _a L_b + L_a \nabla _b 
\right) Y_{\ell m}
(\hat {\boldsymbol{\Omega }}).
\end{aligned}
\end{equation}
Here $a,b=1,2$ are indices of an orthonormal basis on the 2-sphere, 
$\nabla _a$ is the covariant derivative operator on the sphere, and $L_a$ is the angular
momentum operator.   
Polarization is represented as a second-rank traceless symmetric tensor on the celestial
sphere, or a function \smash{$P_{ab}(\hat {\boldsymbol{\Omega }}).$} In a particular direction 
\smash{$\hat {\boldsymbol{\Omega }},$} $P_{ab}$ has two orthogonal principal axes with eigenvalues
equal in magnitude but opposite in sign. The axis of the positive eigenvalues corresponds
to the orientation of a linear polarizer in which the observed CMB temperature 
anisotropy is maximal. 

If we stick to inflation as a explanation of how the primordial cosmological perturbations
were excited, there are two classes of modes excited: (1) the `scalar' modes, which in three dimensions
are represented by a scalar field, and these are linked to the vacuum fluctuations of the 
scalar inflaton field, and (2) the `tensor' perturbations, 
or gravitational waves,
which correspond to the transverse traceless
component of the space-space part of the metric perturbation $h^{ij},$ which is also excited
during inflation on account of the Heisenberg uncertainty relation. 
The former 
are sensitive to a combination of the height and slope of the inflationary potential
whereas the latter are sensitive only to the height of the inflationary potential. 

In terms of the CMB, at linear order `scalar' modes are unable to excite
$B$ modes. Only `tensor' (and possibly `vector') modes are able to excite the $B$ mode polarization
anisotropies, whereas
both three-dimensional `scalar' and `tensor' modes are able to excite $E$ mode polarization anisotropies. 
In Fig.~\ref{AllSpectraPlot} are plotted the CMB anisotropies predicted for the scalar and tensor modes from
inflation. For the scalar modes we use the `concordance' values of the cosmological parameters.
For the tensor modes, from theory we know the spectral index of the tensor modes (accurately
enough for our purposes) but we do not know their amplitude, which is a free parameter on which
presently there are only upper bounds. There has not yet been a first detection (although there 
has been a false alarm as detailed in the following section). The $B$ mode amplitude is typically
characterized in terms of $r$ or $T/S,$ which is the tensor-to-scalar ratio (at 
a particular reference comoving wavenumber). On the plot three possible values are
indicated, $10^{-1},$ $10^{-2},$ and $10^{-3}$ (the highest of which has already been ruled out). 
On the plot the predicted curves rigidly slide up and down as $r$ is varied given the logarithmic vertical
scale.

In the early days of CMB observation $r$ was constrained using the shape of the $TT$ 
CMB angular power spectrum. 
The scalar and tensor spectra add in quadrature, so that
\smash{$C_\ell ^{TT\,\textrm{(tot)}} = C_\ell ^{TT\,\textrm{(scl)}} +r\,C_\ell ^{TT\,\textrm{(ten)}}(r\!=\!1).$} 
WMAP and Planck obtained 
$r<0.29$ (assuming $n_S<1$)\cite{wmapFirstYearsSperg}
and 
$r<0.26$ ($r<0.11$ assuming $n_S=1$)\cite{planck2013params}
using the temperature data alone, respectively. But this
approach is limited due to model uncertainty and also because the scalar contribution acts
as a sort of noise limiting the upper bounds that can be established on $r.$ 

Given the 
smallness of $r,$ a better and more powerful approach is to search for primordial
$B$ mode polarization anisotropies, for which there is no primordial scalar background, at least at linear order.
For low values of $r,$ scalar modes 
beyond linear order 
however introduce a background, 
which must 
be taken into account. 
Gravitational lensing of the scalar perturbations, which is a higher order (quadratic) effect, causes
\smash{$C_\ell ^{EE}$} to become partially lensed into \smash{$C_\ell ^{BB}.$}
At low $\ell $, this \smash{$C_\ell ^{BB\,\textrm{(lensed)}}$} contribution has a spectrum which looks like
white noise [i.e., \smash{$C_\ell ^{BB\,\textrm{(lensed)}} \approx \textrm{(constant)}$}] but at higher 
$\ell $ this spectrum turns over, as shown in Fig.~\ref{AllSpectraPlot}.
The two curves almost coincide for
$r\approx 10^{-2}$ as shown in the plot if one ignores the so-called `reionization bump,' 
which boosts the lowest $(\ell \lesssim 7)$ multipoles. 
This enhancement provides a window for detecting primordial $B$ modes using observations on
very large angular scales. The other window for detecting primordial $B$ modes
lies on intermediate scales, around the so-called `recombination bump.' 
As the plot shows, 
\smash{$C_\ell ^{BB\,\textrm{(lensed)}}$ turns over at high-$\ell $} 
and 
\smash{$C_\ell ^{BB\,\textrm{(ten)}}$} turns over at lower $\ell $ (around $\ell \approx 100).$

Gravitational lensing of the CMB is interesting in its own right and provides invaluable
information as to the matter inhomogeneities situated between us and the last scattering surface. 
We first describe the effect of gravitational lensing qualitatively. In Fig.~\ref{fig:PlanckCMB-Only}
for example, we see the CMB primary anisotropies. Most visible to the eye are spots having a 
characteristic certain size corresponding to the first acoustic peak. A scale invariant
pattern on the sky would look quite different. These spots have a certain average size
and also are on the average circular. In any case no particular orientation for their 
ellipticity is preferred. Gravitational lensing has two effects: on the one hand, it
magnifies or demagnifies; one the other hand, it shears, mapping circles into ellipses having
the same area. From such distortions a map
of the most likely gravitational lensing field may be constructed. Such a map
will have errors as we do not know what the unlensed map would look like. 
However for small wavenumbers of the lensing field the statistical error can be quite 
small, especially if the resolution of the CMB map is very high. 
Hu\cite{huLensingA} and Hu and Okamoto\cite{huLensingB}
showed how to reconstruct the lensing potential
using a harmonic approach. 
For a more detailed up-to-date and quantitative discussion, see the review by Lewis and Challinor.
\cite{lewisChallinor}

Lensing of the CMB polarization allows for a better 
reconstruction of the lensing potential because of
the near absence of primordial $B$ modes: observationally
as far as we know today there are no primordial $B$ modes.
In principle, except for a few degenerate modes
if we have a clean map of the $E$ mode and of the $B$ mode
polarized sky the lensing potential can be reconstructed
with no error. As a practical matter, the situation is not
so simple. 

%
%

Gravitational lensing is an impediment to detection of primordial $B$ modes. 
It acts much like just another background noise, albeit one that does not 
diminish with increased integration time. In Fig.~\ref{AllSpectraPlot}
we see that for $r\approx 10^{-2},$ at intermediate $\ell ,$
between the `reionization bump' and where the primordial $B$ mode signal
starts to plummet (at around $\ell \approx 10^2$), the gravitational lensing and 
primordial contributions to the $B$ mode
signal are of approximately the same magnitude. There are two approaches
to dealing with this interference from gravitational lensing: (1) to simply
treat it as a noise, whose shape and amplitude have been determined from
the fits to the other CMB anisotropies (and also to the $B$ mode signal at much
higher $\ell $), or (2) to try at least partially to subtract the 
contribution from gravitational lensing
at low $\ell $ using $E$ mode and $B$ mode data on smaller angular scales
to reconstruct the gravitational lensing potential, which can
be combined with the $E$ mode data on larger angular scales to obtain a prediction
for the $B$ mode contribution from lensing on large scales, which can
be subtracted from the observed $B$ mode data to obtain a
`delensed' $B$ mode map. For delensing to be effectove, polarization data with very low
noise extending to very small angular scales is required. A detailed
quantitative discussion of this technique can be found in 
Ref.~\citen{coreLensing}. 


Another area of active current discussion is whether the current CMB data 
provides statistically significant evidence that 
the cross-correlation  power spectrum
$C_\ell ^{EB}$ is nonzero.
If the physics responsible for imprinting the primordial cosmological perturbations
and the physics responsible for the propagation of the CMB photons from
the last scattering surface to us today respect parity symmetry (i.e., spatial
inversion symmetry), then $C_{\ell \,\textrm{th}} ^{EB}$ is equal to zero exactly,
because
$a_{\ell m}^E$
and
$a_{\ell m}^B$
acquire opposite signs under spatial inversion.
But $C_{\ell \,\textrm{th}} ^{EB}$ is the average over an infinite number
of sky realizations, whereas $C_{\ell \,\textrm{obs}} ^{EB}$ is calculated
over just a single sky realization and thus is subject to cosmic variance.
Consequently 
$C_{\ell \,\textrm{obs}} ^{EB}$ will be nonzero, and one must determine
whether there is statistically significant nonzero signal above the expected cosmic variance
noise. In certain theories with `cosmic birefringence,' the direction of the linear
polarization of a photon twists in a frequency independent manner as it propagates.
(This is much like Faraday rotation in a magnetized plasma where the linear polarization
twists in a manner proportional to the electron density, the component of the magnetic
field along the direction of propagation, and the square of the wavelength. 
But for cosmic birefringence, 
no plamsma, nor magnetic field is needed, nor is the effect dependent on wavelength.)
In Ref.~\citen{eskilt} an analysis of the WMAP and Planck data suggests a signal
at modest statistical significance. One of the challenges with this type of
measurement is calibrating the polarization orientation of the detectors.

\section{$B$ Mode Searches} 

In 2014 the BICEP2 Collaboration using
a telescope at the South Pole
announced a first detection of
primordial $B$ modes based on
observations in a single
150 GHz frequency channel
claiming a value of
$r=0.20^{+0.07}_{-0.05}$ with $r=0$ disfavored
at $7\sigma .$\cite{bicep2originalPreprint,bicepPublished}
A subsequent analysis 
from Planck
using polarized maps of the same sky region at higher frequencies, 
however, demonstrated that
the signal seen
by BICEP2 in their 150 GHz map was
entirely consistent with polarized
dust emission from our own
Galaxy,\cite{planckXXX}
and a subsequent
joint Plank/BICEP2 analysis established
an upper limit of 
$r<0.12$ at 95\% confidence.\cite{2015PhRvL.114j1301B}
(See also Ref.~\citen{flauger}.)
Since that time upper
limits on $r$ have 
improved.


The problem with this purported detection was a misestimation 
of the possible contribution of polarized dust emission 
to the 150 GHz sky map. A cursory inspection of the CMB sky maps
at different frequencies reveals strong emission from our Galaxy,
especially at very low frequencies and very high frequencies. 
At low frequencies synchrotron emission resulting from
high-energy electrons spiralling in the magnetic field
of the Galaxy dominates. At high frequencies so-called `thermal' 
dust emission dominates. These maps are plotted in Galactic
coordinates using a Mollweide projection, so that the Galactic
center is in the middle and the Galactic plane corresponds to
the equator. Fig.~\ref{fig:PlanckTmp} shows the Planck temperature maps 
and Fig.~\ref{fig:PlanckPol} shows
the Planck polarization maps. Because the degree of polarization 
of the primordial CMB is smaller than that of the Galactic foregrounds,
the width of the band about the Galactic plane appears even wider in 
the polarized maps. 

At the time of the purported BICEP2 detection, the Planck collaboration 
had not released its polarization data. The Planck 2013 release was 
based only on the temperature data, and it was not until
the Planck 2015 release that polarization results were reported. 
It is challenging to measure the thermal dust emission from
the ground because of atmospheric emission which increases with
frequency. Bands must be carefully chosen to avoid particular 
lines of strong atmospheric emission. 

Unlike space-based experiments where there is no room for periodic
upgrades, ground-based CMB experiments periodically upgrade and 
try out various improvements. While it was widely recognized that a convincing
$B$ mode detection would require multifrequency maps in order
to remove foreground contaminants and make a convincing case
for the primordial origin of the observed $B$ mode signal, 
the strategy at the time was to debug at low frequency, 
postponing the addition of higher frequency channels until
a later stage.

\begin{table}
\tbl{\textbf{\boldmath Upper Bounds on Tensor-to-Scalar Ratio $r\!=\!T/S$.}
{\rm Here $r$ is defined as the ratio of the primordial amplitudes at
the pivot wavenumber $k_\textrm{piv} =0.05\, \textrm{Mpc}^{-1}.$}
}
{\begin{tabular}{|l|l|}
\hline
\hline 
Experiment  & Upper Bound (95\% Confidence) \\
\hline
BICEP2/Keck and Planck Collaborations 
\cite{2015PhRvL.114j1301B} & $r < 0.12$ \\
BICEP2/Keck (2018)  \cite{2018PhRvL.121v1301B} & $r<0.07$ ($r<0.06$ with Planck data) \\
Planck NPIPE (2020) \cite{Tristram:2020wbi}
& $r<0.056$ \\
BICEP/Keck (2021) 
\cite{2021PhRvL.127o1301A} &
$r< 0.036$ \\
\hline
\end{tabular}}
\label{Table:ToverSupperLimitss}
\end{table}

At the time Planck did have a chance to detect primordial $B$ 
modes provided that their amplitude, characterized by the 
primordial tensor-to-scalar ratio, conventionally denoted 
as $r$ or $T/S,$ was sufficiently large. 
Efstathiou and Gratton \cite{efstathiou2009b}
for example in a perhaps optimistic
forecast aimed at justifying to ESA an extension 
of the Planck mission 
beyond the nominal 14-month mission\footnote{%
The final Planck survey ended after 30 months
of data taking when the ${}^3He$-${}^4He$ coolant ran out.}
forecast that Planck would be able to
make a reliable detection of
$B$ modes for $r>0.05.$
Since the BICEP2 claim,  observational constraints on $r$ 
have greatly improved, as shown in Table \ref{Table:ToverSupperLimitss}.
In the next section more details of current efforts to detect primordial
$B$ modes are described. 

\section{Future Prospects}

Beyond increasing the precision of tests of the so-called 
`standard cosmological model,' or concordance model,
a number of important big questions remain: 
(1) detection of primordial $B$ modes, 
and (2) precision characterization of the gravitational lensing of the CMB. 

It is useful to compare what has been accomplished so far with the cosmic variance limit resulting 
from the impossibility of observing more than a single sky as already discussed above
[see eqns.~(\ref{CVeqnFirst})--(\ref{CVeqnLast})]. 
For the $\smash{C_\ell ^{TT}},$ $\smash{C_\ell ^{TE}},$ and $\smash{C_\ell ^{EE}},$
the present observations are close to the cosmic variance limit in the multipole range
where the primordial anisotropies dominate, so little is to be gained through better observations,
although the low-$\ell $ polarization observations are dominated by systematic errors, so there is scope
for some improvement. The primary area where great improvement is possible is better observations 
of the $B$ mode polarization anisotropy power spectrum. Sky maps of the $B$ polarization are also of great
interest because these can be used to reconstruct the CMB gravitational lensing potential. CMB lensing 
maps can be cross-correlated with other data, for example from galaxy surveys, allowing biases to be
determined.\cite{coreLensing} 
CMB lensing can be used to constrain absolute neutrino masses because massless neutrinos
lead to free streaming on small scales changing the shape of the lensing spectrum. Neutrino masses however
prevent such free streaming. 

This section covers both experiments in the process of being deployed and experiments still in the planning 
stage. The description here is tentative as the funding situation, especially in the US, has become 
complicated and many of the details in this section are likely to become quickly out of date. Nevertheless,
no matter how things evolve, the design studies and science forecasts cited in this section will continue
to be relevant and serve as the basis for a next generation of CMB probes. 

Today CMB observations have become hard. In the 1990s a plethora of small experiments 
from the ground and from stratospheric balloons\footnote{%
The site https://lambda.gsfc.nasa.gov/papers/ provides an almost comprehensive listing of CMB experiments.}
followed the 1992 COBE DMR CMB anisotropy detection. Through these efforts
experimental techniques were perfected. However today the cost of carrying out a competitive CMB
experiment has skyrocketed. Consequently groups have combined forces to carry out
a smaller number of experiments involving large collaborations, in most cases 
international in character.
These efforts divide into two classes: (1) space-based experiments with small aperture telescopes
taking advantages of the lack of atmospheric interference and incredibly stable observing 
conditions in space, but suffering from low-angular resolution and also limited bandwidth 
in sending data back to earth; and (2) ground-based experiments, sited at locations 
on earth with exquisite observing conditions (for the most part either in the desert in 
northern Chile or at the South Pole), often having large apertures (up to $\approx 10$m)
and thus exquisite angular resolution. 

Although there is a loophole, the conventional wisdom is that the only way to increase sensitivity
substantially is to deploy more detectors. Most CMB telescopes use what is known as single-mode detection,
meaning that each detector receives energy only from a single transverse mode of the incoming radiation
field. Although this was hardly the case earlier, modern bolometric detectors add an amount of noise 
comparable to the intrinsic photon counting noise of the incoming radiation from the sky. This means 
that there is marginal benefit to developing less noisy detectors and the only way to
increase sensitivity is to deploy more detectors. 
This is not quite the whole story because multimode detection is possible, but this comes at the expense
of angular resolution. Unlike the COBE DMR, the COBE FIRAS instrument used two bolometric detectors
capturing a large number of transverse modes, and the PIXIE proposal (which was not ultimately
successful) proposed to detect the polarization anisotropy on large angular scales (in order to detect
$B$ modes) in addition to improving on the FIRAS constraints by several orders of magnitude. 

From space, probably the most likely CMB mission to go forward is the Japanese LiteBIRD mission, which
has been in the planning stage and may be rescoped. The proposed LiteBird mission is described
in detail in Refs.~\citen{liteBirdPTEPreview} and \citen{liteBirdDesign}. 
Another future experiment from the ground for which substantial development and planning
has taken place is CMB Stage-4 (or CMB-S4). 
\cite{cmbS4-constrFore,cmbS4-decadalRev}.
Unfortunately, this US-led experiment, to be funded jointly
by the DOE and the NSF and envisaging some international participation, was cancelled. It is
however conceivable that in the future something very similar to CMB-S4 will be reproposed
and perhaps go forward, and in this case much of the planning and studies will be extremely
useful. The CMB-S4 Science Book\cite{cmbS4scienceBook} describes the case for this experiment
and makes projections for the science that can be realized, and the Technology Book provides
a summary of the state of present relevant technology.\cite{cmbS4techBook}
There is also a Reference Design\cite{cmbS4refDesign}. 
The Snowmass Neutrino document\cite{snowmassNuetrino} explains the role CMB observations can 
play in fixing absolute neutrino masses. (See also for example Ref.~\citen{errard} and 
references therein for a discussion of some of the practicalities of $B$ mode measurements.)

In medium term the Simons Observatory will provide ground-based observations well beyond the capacities
of previous facilities. We may hope that LiteBird moves forward providing $B$ mode observations on
large and intermediate angular scales with the possibility of detecting primordial $B$ modes. 
Activity based at the South Pole also continues.  How
to perform more ambitious observations has been extensively investigated but the funding situation remains
uncertain. 

Another promising area of investigation is measuring the CMB frequency spectrum with greater sensitivity.
Since the COBE DMR discovery of CMB temperature anisotropy, tremendous progress has been made in mapping 
out the CMB temperature and polarization anisotropies. However the COBE FIRAS measurement almost 35 years later
still stands as the best measurement of the CMB frequency spectrum. Ref.~\citen{chluba} reviews what new science would result 
from improved measurements of the CMB frequency spectrum. Several proposals are underway for a follow-up mission
to COBE FIRAS. The Pixie proposal,\cite{pixie1} which ultimately was not successful, describes in detail how
one can improve on FIRAS by a few orders of magnitude.

\vskip 10pt 

\noindent
\textbf{Acknowledgements:} The author thanks Josquin Errard and Ken Ganga for useful comments. 


\end{document}